\documentclass[%
 aip,
 cha,
 amsmath,amssymb,
 reprint,%
]{revtex4-1}

\usepackage{graphicx}
\usepackage{dcolumn}
\usepackage{bm}
\usepackage[mathlines]{lineno}

\usepackage[utf8]{inputenc}
\usepackage[T1]{fontenc}
\usepackage{hyperref}
\usepackage{mathptmx}
\usepackage{mathtools}
\usepackage{etoolbox}
\usepackage{xcolor}
\usepackage{soul}
\usepackage{url}

\makeatletter
\def\@email#1#2{%
 \endgroup
 \patchcmd{\titleblock@produce}
  {\frontmatter@RRAPformat}
  {\frontmatter@RRAPformat{\produce@RRAP{*#1\href{mailto:#2}{#2}}}\frontmatter@RRAPformat}
  {}{}
}%
\makeatother

\begin{document}

\preprint{AIP/123-QED}

\title[Incorporating Coupling Knowledge into Echo State Networks for Learning Spatiotemporally Chaotic Dynamics]{Incorporating Coupling Knowledge into Echo State Networks for Learning Spatiotemporally Chaotic Dynamics}
\author{Kuei-Jan Chu}
    \thanks{Electronic mail: kueijanchu@outlook.com} 
\author{Nozomi Akashi}
    \thanks{Electronic mail: akashi.nozomi.2a@kyoto-u.ac.jp}
\author{Akihiro Yamamoto}%
\affiliation{ 
Graduate School of Informatics, Kyoto University, Kyoto 606-8501, Japan
}%

\date{\today}

\begin{abstract}
Machine learning methods have shown promise in learning chaotic dynamical systems, enabling model-free short-term prediction and attractor reconstruction. However, when applied to large-scale, spatiotemporally chaotic systems, purely data-driven machine learning methods often suffer from inefficiencies, as they require a large learning model size and a massive amount of training data to achieve acceptable performance. To address this challenge, we incorporate the spatial coupling structure of the target system as an inductive bias in the network design. Specifically, we introduce physics-guided clustered echo state networks, leveraging the efficiency of the echo state networks as a base model. Experimental results on benchmark chaotic systems demonstrate that our physics-informed method outperforms existing echo state network models in learning the target chaotic systems. Additionally, we numerically demonstrate that leveraging coupling knowledge into ESN models can enhance their robustness to variations
of training and target system conditions. We further show that our proposed model remains effective even when the coupling knowledge is imperfect or extracted directly from time series data. We believe this approach has the potential to enhance other machine-learning methods.
\end{abstract}

\maketitle

\begin{quotation}
Chaotic systems are ubiquitous in nature and society. Many such systems, such as those found in the brain and social networks, have governing equations that remain largely unknown. This necessitates data-driven approaches to reconstruct their dynamics from observed time-series data. However, for large-scale systems, purely data-driven methods often require massive amounts of training data and computational resources, highlighting the need to integrate system-specific knowledge into data-driven methods. In this work, we incorporate coupling knowledge from target systems into model design. Our tests suggest that the coupling structure serves as an effective inductive bias for learning spatiotemporally chaotic dynamical systems.
\end{quotation}

\section{\label{sec:Intro}Introduction}
Complex dynamical systems (DSs), which span a wide range of spatial and temporal scales, play a crucial role in various scientific and engineering disciplines, including meteorology,~\cite{Kashinath2021PhysicsinformedML} fluid dynamics,~\cite{wang2020towards} and cosmology.~\cite{bohmer2017dynamical} In many cases, such as brain activity and social networks, the governing equations of these systems remain unknown. Understanding the dynamics of chaotic systems is crucial for predicting their behavior, uncovering hidden governing principles, and harnessing their complexity for practical applications in science and technology. These reasons necessitate data-driven approaches to infer their dynamics from observed (ground truth) time-series data. 
Furthermore, most complex DSs exhibit inherent chaos characterized by the exponential divergence of nearby trajectories,~\cite{strogatz2024nonlinear} making them challenging to accurately simulate using model-based classical numerical methods with the governing equation. This challenge is further exacerbated in data-driven approaches designed to forecast the system’s future states. Due to the inherent chaoticity in those complex DSs, our focus in the context of DS learning shifts from state forecasting to attractor reconstruction.
This entails learning a generative model capable of replicating the system’s invariant (long-term) geometrical and temporal properties.~\cite{Hess2023GeneralizedTF}

Given these challenges, machine learning (ML) methods—particularly recurrent neural networks (RNNs)—have emerged as powerful data-driven approaches for learning chaotic DSs in recent years.~\cite{jaeger2001echo, pathak2018model, lu2018attractor} However, for complex, high-dimensional DSs, purely data-driven learning requires an enormous amount of training data, which is often impractical due to the high cost and difficulty of data collection in many scientific fields.~\cite{karniadakis2021physics} Additionally, RNN-based methods typically rely on backpropagation through time training,~\cite{werbos1990backpropagation} which introduces inefficiencies during the unrolling of the computational graph and suffers from vanishing and exploding gradients.~\cite{vlachas2020backpropagation} Moreover, compared to popular but complex gated RNN models, such as long short-term memory (LSTM)~\cite{hochreiter1998vanishing} and gated recurrent unit (GRU),~\cite{cho2014learning} simpler architectures offer advantages in mathematical tractability and scientific interpretability for learning chaotic systems.~\cite{Hess2023GeneralizedTF} All these reasons highlight the need to integrate system-specific knowledge into the learning process using a simple ML architecture that balances efficiency and interpretability.~\cite{karniadakis2021physics, pathak2018model, Pathak2018HybridFO}

Among the various forms of system-specific knowledge that can aid DS learning through ML, physical information—such as constraints based on physical models—has gained increasing attention and demonstrated significant potential in recent studies.~\cite{wang2020towards, pathak2018model, arcomano2022hybrid, Pathak2018HybridFO, Darbon2020OnSN, raissi2019physics, sun2021physics, sun2022self, sun2022physics}
These studies have integrated prior physical knowledge into neural network design by structuring model components accordingly,~\cite{wang2020towards, pathak2018model, sun2021physics} embedding mathematical solutions directly,~\cite{Pathak2018HybridFO, arcomano2022hybrid, Darbon2020OnSN, sun2022self, sun2022physics} or formulating loss functions based on the residual of the governing partial differential equations (PDEs).~\cite{raissi2019physics} 
For example, Arcomano et al.~\cite{arcomano2022hybrid} proposed a computationally efficient hybrid model that combines a physics-based global atmospheric circulation model with a machine learning component for weather forecasting, achieving improved quantitative forecasting performance.


Building on the above studies, physical knowledge is also expected to contribute to DS learning as a inductive bias for designing specialized structures inside a network layer of an ML model.
In this work, we adopt an Echo State Network (ESN)~\cite{jaeger2001echo} as the base model for DS learning. The ESN, an RNN model within the reservoir computing~\cite{verstraeten2007experimental} paradigm, is well-suited for sequential data processing, particularly in attractor reconstruction, due to its echo state property and efficient training mechanism.~\cite{Pathak2017UsingML, jaeger2001echo} Inspired by the work of Ref.~\onlinecite{pathak2018model}, which introduced a parallelized ESN structure leveraging the local interactions of the target DS in the input layer, we propose a novel model called the physics-guided clustered ESN (PGC-ESN). Our approach leverages the spatial coupling structure of the target DS as prior physical knowledge to design a coupled clustered architecture in the ESN's reservoir layer, which typically has random connectivity. 
The effectiveness of multiple clustered structures in ESNs has been studied in Ref.~\onlinecite{gallicchio2017deep, kawai2019small, li2022multi}.

In practical scenarios, prior knowledge of the target system’s coupling structure is typically imperfect or even unavailable. 
To simulate the impact of using guidance from imperfect prior knowledge, we evaluate PGC-ESN with a perturbed coupling structure.
Inspired by the work of Ref.~\onlinecite{goldmann2022learn}, which extracted the delay of the Mackey–Glass system from time series data to enhance the predictive ability of a model known as delayed echo state network, we also extract the coupling structure directly from time series data and use this information to guide the design of PGC-ESN.

Empirical evaluations on two benchmark spatiotemporal dynamical systems, the Lorenz-96 system~\cite{lorenz1996predictability} and the Kuramoto–Sivashinsky equation~\cite{kuramoto1976persistent, ashinsky1988nonlinear}, demonstrate that when the target system exhibits a sufficient coupling structure, our proposed method outperforms the standard ESN (Std-ESN), a randomly clustered ESN (RandC-ESN, a non-coupled clustered ESN model we made for comparison), and a parallelized ESN (Paral-ESN) in long-term attractor reconstruction under limited model capacity, and in short-term prediction even when the model size is sufficiently large. Experiments on the Lorenz-96 system with varying training noise level, training data size, sequence temporal resolution, and increasing system nonlinearity further show that the physics-informed methodologies used in PGC-ESN and Paral-ESN can enhance the robustness of ESN models to varied learning settings and improve their learning performance even when the target system's nonlinearity increases. Additionally, we demonstrate the effectiveness of PGC-ESN numerically on the Lorenz-96 system when provided with imperfect prior or extracted coupling knowledge, highlighting the practical applicability of our method in scenarios where prior knowledge of the target system is imperfect or even unavailable.

Overall, we numerically show that the coupling structure inherent in a DS can serve as an inductive bias to guide network design for DS learning. We anticipate that in the future, our work will inspire the development of other artificial neural networks that incorporate system-specific knowledge, not only for DS learning but also for broader ML tasks related to structured data.

The rest of this paper is organized as follows. Section \ref{sec:DS_reconstruction_background} provides background information on attractor reconstruction, including the definition of a DS and a guide to setting up ESNs to carry out this task. Section \ref{sec:Proposed_method} details how we incorporate the coupling structure of a target system as prior or extracted physical knowledge into our proposed method. Section \ref{sec:Experiments} presents the evaluation results. 
Finally, Section \ref{sec:conclusion} concludes the paper with a summary of our contributions and several promising directions for future research.

\section{\label{sec:DS_reconstruction_background}Dynamical system reconstruction background}
\subsection{\label{subsec:dynamical_systems}Dynamical systems}
While both continuous and discrete Dynamical systems exist, we focus here on learning continuous DSs governed by ordinary and partial differential equations related to one or more unknown functions and their derivatives. The following definition is adapted from Ref.~\onlinecite{yu2024learning}.

\paragraph{Definition.}
Let $k \geq 1$ be an integer and $U$ be an open subset of $\mathbb{R}^n$ representing the system domain. We define the system state $\boldsymbol{u} : U \rightarrow \mathbb{R}^m$ 
and a variable $\boldsymbol{x} \in U$. An expression of the form
\begin{equation}
  \mathcal{F}(D^k \boldsymbol{u}(\boldsymbol{x}), D^{k-1} \boldsymbol{u}(\boldsymbol{x}), \dots, D \boldsymbol{u}(\boldsymbol{x}), \boldsymbol{u}(\boldsymbol{x}), \boldsymbol{x}) = \boldsymbol{0}
  \label{eq:DS}
\end{equation}
is called a $k^\text{th}$-order system of partial (or ordinary when $n=1$) differential equations, where $\mathcal{F} : \mathbb{R}^{mn^k} \times \mathbb{R}^{mn^{k-1}} \times \dots \times \mathbb{R}^{mn} \times \mathbb{R}^m \times U \rightarrow \mathbb{R}^m$, and $D$ denotes either a partial or ordinary derivative operator.

The operator $\mathcal{F}$ governs the $m$-dimensional DS with respect to the variable $\boldsymbol{x} \in \mathbb{R}^n$ and can be either linear or non-linear. Nonlinearity is often responsible for the chaotic nature of a DS. Since most DSs evolve over time, and one of the variables in $\boldsymbol{x}$ is the time dimension, we commonly use the shorthand notation $\boldsymbol{u}(t)$ to represent $\boldsymbol{u}(t, x_1, \dots, x_{n-1})$, where $\{ x_1, \dots, x_{n-1} \}$ are the spatial parts of the variable $\boldsymbol{x}$. 

Chaotic DSs exhibit a sensitive dependence on initial conditions, meaning that nearby trajectories diverge exponentially over time, making future state forecasting inherently challenging. This exponential divergence rate is called the maximum Lyapunov exponent (MLE),~\cite{strogatz2024nonlinear} denoted as $\lambda$. Due to their chaotic nature, we focus on reconstructing chaotic DSs in this work, emphasizing the learning of the system's long-term behavior.
For short-term state prediction, we use the Lyapunov time,  defined as $T_{\lambda} = {\lambda}^{-1}$, as a reference timescale to measure the growth of error between predicted and observed sequences.

\subsection{\label{subsec:DS_reconstruction_using_ESNs}Dynamical system reconstruction using echo state networks}
In the task of attractor reconstruction, the objective is to learn a generative model that approximates the underlying dynamics of an unknown system from observed time-series data. This model should be capable of generating sequences that replicate the long-term behavior of the system, including the geometry of its dynamical structures in state space, such as attractors, and invariant temporal properties, such as the system's power spectrum, after sufficient training.~\cite{Hess2023GeneralizedTF} In this context, the goal is to implicitly approximate the operator $\mathcal{F}$ on the attractor, which governs the evolution of the underlying DS, as described in Equation~\ref{eq:DS}. To achieve this using an ML model, we sample an observed time series $ \boldsymbol{u}_{1:T} = \{ \boldsymbol{u}(1) \dots \boldsymbol{u}(T) \}$ of length $T$ from the ground truth system at a fixed integration step size $ \Delta t$, where each $\boldsymbol{u}(t) \in \mathbb{R}^{d_u} $ is the state at time step $t$ in the state space. The training objective here is a one-step prediction: given the current state $\boldsymbol{u}(t)$, the model outputs the prediction $\hat{\boldsymbol{u}}(t+1)$ for the next time step.

An ESN model, as an artificial neural network example of the reservoir computing paradigm, uses a recurrent internal layer called a reservoir, typically coupled with linear input and readout layers.~\cite{jaeger2001echo} The weights in the input and reservoir layers are initialized and remain fixed throughout the learning process, with only the readout weights being trained. This design alleviates the complexities of learning recurrent connections that emerge in other RNN optimization methods, such as backpropagation through time.~\cite{vlachas2020backpropagation}

We provide a schematic illustration of the ESN learning process in Appendix~\ref{Appendix_Learning_process_for_ESNs}, along with the following mathematical explanation for our setup. At time step $t$ of the training phase, the input layer $\hat{\mathbf{R}}_\mathrm{in}$ embeds the input $\boldsymbol{u}(t)$ into the reservoir state $ \boldsymbol{r}(t) \in  \mathbb{R}^{d_r}$ linearly by $\hat{\mathbf{R}}_\mathrm{in} [\boldsymbol{u}(t)] = \boldsymbol{W}^{\mathrm{in}} \boldsymbol{u}(t)$, where $d_r$ is the reservoir size. The readout layer maps the reservoir state to the output $\hat{\boldsymbol{u}}(t+1)$ linearly by $ \hat{\boldsymbol{u}}(t+1) = \hat{\mathbf{R}}_\mathrm{out} [\boldsymbol{r}(t)] = \boldsymbol{W}^{\mathrm{out}} \tilde{\boldsymbol{r}}(t)$. The weight matrices $ \boldsymbol{W}^{\mathrm{in}} \in  \mathbb{R}^{{d_r} \times {d_u}}, \boldsymbol{W} \in  \mathbb{R}^{{d_r} \times {d_r}}$, and $\boldsymbol{W}^{\mathrm{out}} \in  \mathbb{R}^{ {d_u} \times {d_r} }$ respectively. Typically, we set the reservoir layer's dimension ${d_r} \gg {d_u}$ to ensure that it has sufficient capability to capture the temporal features of the observed state history. The augmented reservoir state $ \tilde{\boldsymbol{r}}(t)$ is obtained by squaring half of the elements with even indexes in $\boldsymbol{r}(t)$ to enrich the reservoir dynamics, which empirically improves the model's efficiency in attractor reconstruction.~\cite{pathak2018model}
The entries of $\boldsymbol{W}^{\mathrm{in}}$ are uniformly sampled from $[-\beta, \beta]$ and masked randomly to give the matrix a density $p_{\mathrm{in}}$, where $\beta$ is referred to as the input scaling and controls the relative influence of the current input on the reservoir state versus the history of inputs.~\cite{lukovsevivcius2012practical} The entries of $\boldsymbol{W}$ are uniformly sampled from $[0, 1]$, masked randomly to give this matrix a density $p$, and finally rescaled so that its spectral radius (the largest absolute value of the eigenvalues) $\rho < 1$. This scaling ensures that the ESN satisfies the $ \textit{echo state property}$,~\cite{jaeger2001echo} meaning that the reservoir state $\boldsymbol{r}(t)$ becomes independent of its initial conditions as $t$  increases, given a sufficiently long input sequence. The spectral radius $\rho$ also affects the stability of the reservoir activations and the speed at which the influence of past inputs decays in the reservoir.~\cite{lukovsevivcius2012practical} In our experiments, we tune the spectral radius $\rho$ and the input scaling $\beta$ as hyperparameters. The two densities, $p_{\mathrm{in}}$ and $p$, are fixed when prior coupling knowledge of the target system is available, and are tuned otherwise.

To learn a DS using an ESN, we can begin by initializing the reservoir state to $\boldsymbol{0} \in \mathbb{R}^{d_r}$ and warming it up with an observed sequence of length $T_{\mathrm{warm}}$, both in the training and the prediction phases. This process renders the ESN able to encode temporal dependencies on the past state history into the reservoir state $ \boldsymbol{r}$ and predict multiple sequences without requiring proximity to the training data, highlighting the reusability of the readout layer for generating extended sequences after being trained just once.~\cite{vlachas2020backpropagation} During the training phase, we feed an observed sequence of length $T_{\mathrm{train}}$ into the model and use the open-loop configuration to evolve the reservoir states as follows:
\begin{equation}
  \boldsymbol{r}(t) = f(\boldsymbol{W}^\mathrm{in} \boldsymbol{u}(t) + \boldsymbol{W} \boldsymbol{r}(t-1)),
\end{equation} 
where the activation function $f$ is chosen to be the component-wise sigmoid function $f = \tanh$.  Note that the open-loop configuration is also used during warm-up. We then record the reservoir states of length $T_{\mathrm{train}}$ and train the readout matrix $\boldsymbol{W}^{\mathrm{out}}$ by minimizing the mean squared error (MSE) between the observed and predicted sequences using least-squares regression. To prevent overfitting, a Tikhonov regularization term with parameter $\eta$ is typically added, which penalizes the assignation of large values to the fitting parameters in $\boldsymbol{W}^{\mathrm{out}}$. For further regularization, we add Gaussian noise sampled from a normal distribution $\mathcal{N}\left( \mathbf{0}, \, \kappa^{2} \operatorname{diag}\left(\boldsymbol{\sigma}(\boldsymbol{u}_{1:T_{\mathrm{Warm}}+T_{\mathrm{Train}}})^{2}\right) \right)$ to the observed training sequences $\boldsymbol{u}_{1:T_{\mathrm{Warm}}+T_{\mathrm{Train}}}$, where $\boldsymbol{\sigma}$ denotes the vector of standard deviations over time, computed per state-space dimension of the sequence, and $\kappa$ is the noise level that determines the magnitude of perturbation. To alleviate the RAM requirement and improve the computational efficiency, we use a time-batched schedule, as described in Ref.~\onlinecite{vlachas2020backpropagation}. The loss function for training is as follows:
\begin{equation}
  \sum_{1 \leq t \leq T_{\mathrm{train}}} \lVert  \hat{\boldsymbol{u}}(t) - \boldsymbol{u}(t) \rVert^2 + \eta \lVert \boldsymbol{W}^{\mathrm{out}} \rVert^2.
\end{equation}

Once sufficiently trained, we switch to the prediction phase. The reservoir state evolves autonomously to generate a predicted sequence of length $T_{\mathrm{pred}}$ using the closed-loop configuration as follows:
\begin{equation}
  \boldsymbol{r}(t) = f(\boldsymbol{W}^\mathrm{in}  \hat{\boldsymbol{u}}(t) + \boldsymbol{W} \boldsymbol{r}(t-1)),
\end{equation} 
where $\hat{\boldsymbol{u}}(t) = \boldsymbol{W}^{\mathrm{out}} \tilde{\boldsymbol{r}}(t-1)$ is the predicted state at time step $t$, generated by the ESN using the trained $\boldsymbol{W}^{\mathrm{out}}$.

\subsection{\label{sec:Parallelized_esn}Parallelized echo state network}
Pathak et al.~\cite{pathak2018model} proposed a learning system that leverages the local interactions within an observed PDE system—specifically, the Kuramoto–Sivashinsky equation—to design a parallelized reservoir computing framework. In this approach, multiple reservoir computers operate in parallel, each receiving from the input state only a subset of dimensions with size $G$, along with interacting components with size $I$ on both the left and right sides. Each reservoir computer is trained independently to predict the future state of its assigned dimensions. 
We refer to this approach as the Parallelized ESN (Paral-ESN) in this work. 
The empirical performance of their method in predicting the target system proved superior to that of a standard ESN. 

\begin{figure*}[htbp]  
    \centering
    \includegraphics[width=1.0\textwidth]{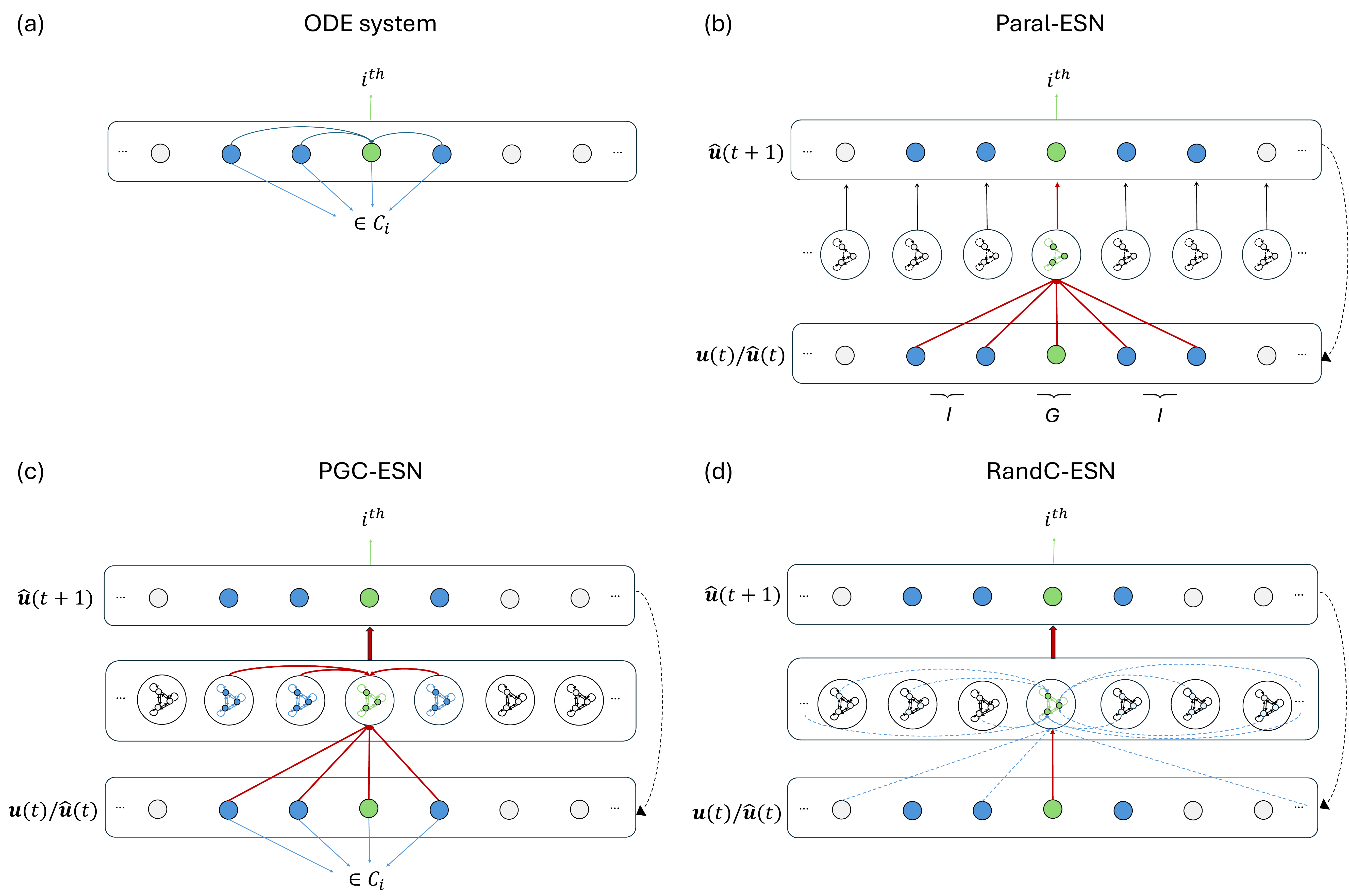}  
    \caption{Illustrative diagram of an example target ODE system (a) and the learning process using Paral-ESN, PGC-ESN, and RandC-ESN in (b), (c) and (d), respectively.}
    \label{fig:ODE_and_ESN_models}
\end{figure*}

Fig.\ref{fig:ODE_and_ESN_models}(a) illustrates an ordinary differential equation (ODE) system, which we will discuss further in Sec.~\ref{subsec:Physical_guidance_in_Dynamical_Systems}, while Fig.~\ref{fig:ODE_and_ESN_models}(b) depicts the corresponding Paral-ESN designed for learning its dynamics. We highlight the $i^{\mathrm{th}}$ reservoir computer and the corresponding dimensions of its input and predicted state with size $G=1$ in green, while the interacting dimensions of the input state with size $I=2$ are colored in blue. Note that the input state may consist of predicted states generated by the ESNs during the automatic prediction phase, where the dashed arrow identifies the feedback connection. This design pattern remains consistent across all reservoirs and is omitted here for brevity.

However, in Paral-ESN, local interactions are incorporated symmetrically only through the input layer. Moreover, ODE systems exhibit more substantial coupling structures, which can also serve as valuable guidance for model design. Building on this idea, we propose an ESN model that leverages physical knowledge of the coupling structure in an ODE system or the local interactions in a PDE system to guide the design of both the input and reservoir layers.

\section{\label{sec:Proposed_method}Proposed method}
\subsection{\label{subsec:Physical_guidance_in_Dynamical_Systems}Physical guidance in Dynamical Systems}
Consider a DS governed by a set of ODEs with system state $\boldsymbol{u} \in  \mathbb{R}^{d_u}$ and a variable $\boldsymbol{x} \in  \mathbb{R}^{n}$. We define the $i^\mathrm{th}$ dimension $u_i$ of the system state as coupled with the vector $\boldsymbol{u}_{c_{i}} = [u_i, u_{c_{i_{1}}}, \dots, u_{c_{i_{l}}} ] \in \mathbb{R}^{l+1} $, supposing that the evolution of $u_i$ is governed by a specific form of Equation $\ref{eq:DS}$ as follows:
\begin{equation}
\mathcal{F}_i (D^k \boldsymbol{u}_{c_{i}}(\boldsymbol{x}), D^{k-1} \boldsymbol{u}_{c_{i}}(\boldsymbol{x}), \dots, D \boldsymbol{u}_{c_{i}}(\boldsymbol{x}), \boldsymbol{u}_{c_{i}}(\boldsymbol{x}), \boldsymbol{x} ) = 0,
\end{equation}
where $D$ represents the ordinary derivative operator, $k$ is the order of the system, the index set $C_{i} = \{ i, c_{i_{1}}, \dots, c_{i_{l}} \}$ is a subset of $\{ 1, \dots, d_u \}$, and $l$ equals the number of its coupled dimensions besides itself, which could be $0$. Fig.~\ref{fig:ODE_and_ESN_models}(a) schematically illustrates an observed ODE system with a coupled structure, which corresponds to the structure of the Lorenz-96 system—a benchmark DS used in this work. As the figure shows, the specific $i^{\mathrm{th}}$ dimension highlighted in green is coupled with its two preceding and one succeeding dimensions, which are colored in blue. Therefore, the index set for the coupled dimensions of the $i^{\mathrm{th}}$ dimension is given by $C_i = \{ i-2, i-1, i, i+1 \}$. The coupling for all other dimensions of this example system is identical and omitted here for brevity.

For DS governed by PDEs, we consider the case where both the system state $\boldsymbol{u}$ and the spatial part of the variable $\boldsymbol{x}$ are one-dimensional. In this scenario, the governing equation takes the following form:
\begin{equation}
\mathcal{F} (D^k u(t,x), D^{k-1} u(t,x), \dots, D u(t), u(t,x), (t,x) ) = 0,
\end{equation}
where $D$ now represents the partial derivative operator. To simulate time-series data from this system, we discretize the spatial variable $x$ into $d_u$ dimensions, yielding a multivariate time series with an observable state $\tilde{\boldsymbol{u}} \in  \mathbb{R}^{d_u}$. The resulting state $\tilde{\boldsymbol{u}}$ exhibits local interactions, which can be leveraged as physical guidance, similar to the coupling structure in ODE systems. We group the dimensions into units of size $G$ and assume that each unit interacts with $L$ adjacent units on both the left and right sides. A schematic illustration of this structure is provided in Appendix~\ref{sec:appendix_PGC-ESN_for_PDE_system}. For the sake of clarity and consistency, we refer to the local interactions as a coupling structure, and the simulated observable state $\tilde{\boldsymbol{u}}$ as $\boldsymbol{u}$ in a PDE system, acknowledging a minor trade-off in the accuracy of these expressions. 

\subsection{\label{subsec:proposed_model}Proposed model}
In addition to the observed time series, suppose we also have prior knowledge of the coupling structure of the underlying DS. We can leverage this information as guidance to design a specialized ESN model, which we call the physics-guided clustered ESN (PGC-ESN). 
To illustrate the proposed approach, we assume that the underlying DS is governed by the ODE system shown in Fig.~\ref{fig:ODE_and_ESN_models}(a) and schematically illustrate the corresponding PGC-ESN in Fig.~\ref{fig:ODE_and_ESN_models} (c). Specifically, our method involves constructing parallel, fully connected clusters within a single large reservoir. Each cluster corresponds to a particular dimension of the observed system and receives both its own dimension and the coupled dimensions of the input state. Additionally, each cluster is fully connected to the specific clusters whose corresponding dimensions of the observed system are coupled with its own. The $i^{\mathrm{th}}$ dimension of the input and predicted states, along with the corresponding cluster in the reservoir, are highlighted in green, while the coupled dimensions and clusters are shown in blue. Unlike the Paralleled ESN method explored in Ref.~\onlinecite{pathak2018model}, which involves training each ESN separately, our proposed model maintains a single unified reservoir, enabling cohesive learning through a standard fully connected readout layer. This design methodology is consistently extended to PDE-based target systems, as illustrated in Appendix~\ref{sec:appendix_PGC-ESN_for_PDE_system}.

The reservoir state update equation for our proposed PGC-ESN in open-loop mode is given by the following equation:
\begin{equation}
  \boldsymbol{r}_i (t) = f ( \sum_{j \in C_i} \boldsymbol{W}^{\mathrm{in}}_{(i,j)} u_{j}(t) + 
  \sum_{j \in C_i} \boldsymbol{W}_{(i,j)} \boldsymbol{r}_{j} (t-1) ),
\end{equation}
where $\boldsymbol{r}_i (t)$ represents the state vector of the $i^{\mathrm{th}}$ cluster in the reservoir, and $u_{j}(t)$ denotes the $j^{\mathrm{th}}$ dimension of the input state at time $t$. The matrix $\boldsymbol{W}^{\mathrm{in}}_{(i,j)}$ contains the weight entries for connections from the $j^{\mathrm{th}}$ dimension of the input state to the $i^{\mathrm{th}}$ cluster in the reservoir, and $\boldsymbol{W}_{(i,j)}$ contains the weight entries for connections within the $i^{\mathrm{th}}$ cluster (if $i = j$) or from the $j^{\mathrm{th}}$ cluster to the $i^{\mathrm{th}}$ cluster in the reservoir (if $i \neq j$), where $j$ belongs to the set of coupled dimensions $C_i$ for the $i^{\mathrm{th}}$ dimension of the observed system. In closed-loop mode, the update equation remains almost the same, except that the reservoir is driven by the predicted state $\hat{\boldsymbol{u}}(t)$. A schematic illustration of the updation is provided in Appendix~\ref{Appendix_Reservoir_updation_for_PGC_ESNs}.

In practical scenarios, it is often necessary to consider settings where only time series data are available, and no prior knowledge of the underlying system’s coupling structure is provided. To address this, we propose a variant of PGC-ESN, referred to as AutoPGC-ESN, which is automatically guided by the extracted coupling information from the observed time series. Specifically, we estimate a coupling strength matrix $\boldsymbol{C}$, where each element $\boldsymbol{C}_{i,j} \geq 0$ quantifies the inferred coupling strength from the $j^{\mathrm{th}}$ dimension (sender) to the $i^{\mathrm{th}}$ dimension (receiver) of the underlying ODE system for $i \neq j$, and is defined as $0$ for $i = j$. In our experiment, this estimation is performed by applying transfer entropy~\cite{schreiber2000measuring} to the pairwise sequences corresponding to dimensions $i$ and $j$ of the observed time series. Details regarding the definition and computation of transfer entropy are provided in Appendix~\ref{sec:appendix_transfer_entropy_details}.
For each dimension $i$, we define its inferred coupled dimensions set $\hat{C}_i$ as the union of $i$ and the $q$ dimensions with the highest (non-self) inferred coupling strength in the $i^{\mathrm{th}}$ row of the coupling strength matrix $\boldsymbol{C}$. The parameter $q$ is shared across all dimensions and affects the density of weight matrices in both the input and reservoir layers, as both densities now equal the ratio of $(1+q)$ to the dimension number of the underlying system. Using $\hat{C}_i$, we construct the coupling structure for the $i^{\mathrm{th}}$ cluster of AutoPGC-ESN's reservoir in the same manner as in PGC-ESN. 
Notably, as the inferred coupling varies across dimensions, the resulting coupling structure of AutoPGC-ESN may differ across clusters accordingly. For PDE systems, a similar approach can be employed, with coupling defined and inferred at the level of spatial units as described in Sec.~\ref{subsec:Physical_guidance_in_Dynamical_Systems}.

\subsection{\label{subsec:comparison_model}Comparison model without physical guidance}
As an ablation study, we evaluate whether introducing the clustered structure in the reservoir layer—serving as the sole inductive bias—can improve learning performance without incorporating any spatial structure knowledge of the underlying system. To this end, we design a comparison model termed randomly clustered ESN (RandC-ESN). As shown in Fig.~\ref{fig:ODE_and_ESN_models} (d), this model consists of a single reservoir composed of parallel, fully connected clusters. Each cluster corresponds to a specific dimension (or unit) of the observed ODE (or PDE) system and receives both its own dimension (unit) and a selection of randomly chosen dimensions (units) from the input state. Inter-cluster neuron connections are randomly activated without any weight bias, ensuring that no spatial structure knowledge of the underlying system is exploited. We use a designed probability governing the random connections in both the input and reservoir layers to ensure that the overall density of these layers matches that of the PGC-ESN.

\section{\label{sec:Experiments}Experiments}
\subsection{\label{subsec:Benchmark_Dynamical_Systems}Benchmark dynamical systems}
\subsubsection{\label{subsubsec:Lorenz_96}Lorenz-96 system}
The Lorenz-96 system~\cite{lorenz1996predictability} is a high-dimensional spatiotemporal DS governed by the following equation:
\begin{equation}
\frac{d\boldsymbol{u}_i}{dt} = \alpha (\boldsymbol{u}_{i+1} - \boldsymbol{u}_{i-2})  \boldsymbol{u}_{i-1} - \boldsymbol{u}_i + F ,
\end{equation}
for $i = 1, \dots, d_{u}$, where the dimension $d_{u} \geq 4$, and $\alpha \geq 0$.
The beginning and end are wrapped around such that $\boldsymbol{u}_{-1} = \boldsymbol{u}_{d_{u}-1}$, $\boldsymbol{u}_0 = \boldsymbol{u}_{d_{u}} $ and $\boldsymbol{u}_{d_{u}+1} = \boldsymbol{u}_1$. This system consists of a nonlinear advective term $\alpha (\boldsymbol{u}_{i+1} - \boldsymbol{u}_{i-2})  \boldsymbol{u}_{i-1}$, a linear damping term $- \boldsymbol{u}_i$, and an external forcing term $F$.  The parameter $\alpha$ is added to control the nonlinearity of the system. When $\alpha = 1$, the system is reduced to the standard Lorenz-96 system. In this work, we set $d_{u} = 40$ and $F = 8$.
To simulate an observed trajectory, we start with a random initial state $\boldsymbol{u}(0) \sim \mathcal{N}(\boldsymbol{F}, \, (0.01)^2 \mathbf{I})$, where $\boldsymbol{F} \in \mathbb{R}^{d_u}$ is a vector with all elements equal to $ F$, and $\mathbf{I} \in \mathbb{R}^{d_u \times d_u}$ is the identity matrix. The system is then integrated using the 4th-order Runge-Kutta method~\cite{butcher2016numerical} with step size $\Delta t = 0.01$. We generate a trajectory of length $T = 2 \cdot 10^5$ by discarding the first $10^5$ samples to eliminate the effects of initial transients.
The MLE for this system was calculated to be $\lambda \approx 1.69$ using the Shimoda-Nagashima method,~\cite{shimada1979numerical} and the corresponding Lyapunov time $T^{\lambda} \approx 0.59$.

\subsubsection{\label{subsubsec:Kuramoto–Sivashinsky}Kuramoto–Sivashinsky equation}
The Kuramoto–Sivashinsky (KS) equation~\cite{kuramoto1976persistent, ashinsky1988nonlinear} is a nonlinear partial differential equation of the fourth order. In this work, we employ a modified version of the one-dimensional KS equation as described in Ref.~\onlinecite{Pathak2017UsingML}, which is given by
\begin{equation}
  \frac{\partial u}{\partial t} = - \frac{\partial^4 u}{\partial x^4} - 
  \left[ 1 + \mu \cos\left(\frac{2\pi x}{\gamma}\right) \right] \frac{\partial^2 u}{\partial x^2} - 
  u \frac{\partial u}{\partial x},
\end{equation}
where $u$ is defined on the domain $ x \in [0, L]$ with periodic boundary conditions red{\st{, that is, }} $u(0,t) = u(L,t)$. 
The equation is spatially inhomogeneous when $\mu \neq 0$ and homogeneous when $\mu = 0$, with both cases being considered in our experiments. The boundary size $L$ controls the dimensionality of the system’s attractor, scaling it linearly when $L$  is large. The system's domain is spatially discretized to $d_u$ nodes with a fixed grid size, yielding an observable state $\boldsymbol{u} \in \mathbb{R}^{d_u}$. We solve the equation using the Exponential Time-Differencing Fourth-Order Runge–Kutta algorithm (ETDRK4) for stiff PDEs,~\cite{kassam2005fourth} with step size $\Delta t = 0.25$. The simulated trajectory has a length of $T = 2 \cdot 10^5$, with the first $4 \cdot 10^4$ samples being discarded to avoid initial transients. In our experiment, we set $L = 100$, $d_u = 256$ for both cases, and $\mu = 0.1$, $\gamma = 25$ for the inhomogeneous case. The calculated MLE values are $\lambda \approx 0.09$ for the homogeneous case and $\lambda \approx 0.08$ for the inhomogeneous case, with corresponding Lyapunov times of $T^{\lambda} \approx 11.02$ and $T^{\lambda} \approx 12.57$, respectively.

\subsection{\label{subsec:Evaluation_metrics}Evaluation metrics}
In this work, we evaluate both geometric and temporal reconstruction using two metrics, referred to as the geometrical distance $D_{\mathrm{geom}}$ and the temporal distance $D_{\mathrm{temp}}$,~\cite{Hess2023GeneralizedTF} for long-term attractor reconstruction, where smaller values indicate better reconstruction. Our aim was to capture the invariant geometrical structure of the system’s attractor in the state space and its temporal signature called the power spectrum. We provide the definitions of the two measures below, with further details on their numerical approximation in Appendix~\ref{sec:appendix_evaluation_metrics_details}. Additionally, we assess the learning system's short-term predictive performance using a metric called valid prediction time (VPT).~\cite{vlachas2020backpropagation} 

\subsubsection{\label{subsubsec:Geometrical_distance}Geometrical distance}
Geometrical distance~\cite{Hess2023GeneralizedTF} is an invariant statistic used to measure the degree of overlap between true and reconstructed attractor geometries. Given the observed and predicted sequences of length $T_{\mathrm{pred}}$ on the invariant attractor, $\{ \boldsymbol{u}(1), \dots \boldsymbol{u}(T_{\mathrm{pred}}) \}$ and $\{ \hat{\boldsymbol{u}}(1), \dots \hat{\boldsymbol{u}}(T_{\mathrm{pred}}) \}$, which are generated by the ground truth system and the reconstructed system (i.e., the trained ML model), we can estimate the corresponding ideal invariant distributions $p(\boldsymbol{u})$ and $p(\boldsymbol{u} | \boldsymbol{\theta})$ over the state space, where $\boldsymbol{\theta}$ denotes the learned model parameters. The geometrical distance $D_{\mathrm{geom}}$ is then defined as
\begin{equation}
  D_{\mathrm{geom}} \coloneqq  D_{\mathrm{KL}} (p(\boldsymbol{u}) \Vert p(\boldsymbol{u} | \boldsymbol{\theta})) = \int_{\boldsymbol{u} \in \mathbb{R}^{d_u}} p(\boldsymbol{u}) \log \frac{p(\boldsymbol{u})}{p(\boldsymbol{u} | \boldsymbol{\theta})} \, d\boldsymbol{u} ,
\end{equation}
where $D_{\mathrm{KL}}$ represents the Kullback–Leibler (KL) divergence between the two distributions. 
Fig.~\ref{fig:Lorenz96_reservoir_size_500_plots}(a) shows both the target and predicted attractors projected onto the first three-dimensional subspace in the evaluation of ESN models on the Lorenz-96 system. The $D_{\mathrm{geom}}$ values reported in Table~\ref{tab:Lorenz96_performance} quantify the discrepancy between the target and predicted attractors for each model, averaged across sequences and model initializations.

\subsubsection{\label{subsubsec:Temporal_distance}Temporal distance}
\begin{figure*}[htbp]  
    \centering
    \includegraphics[width=1\textwidth]{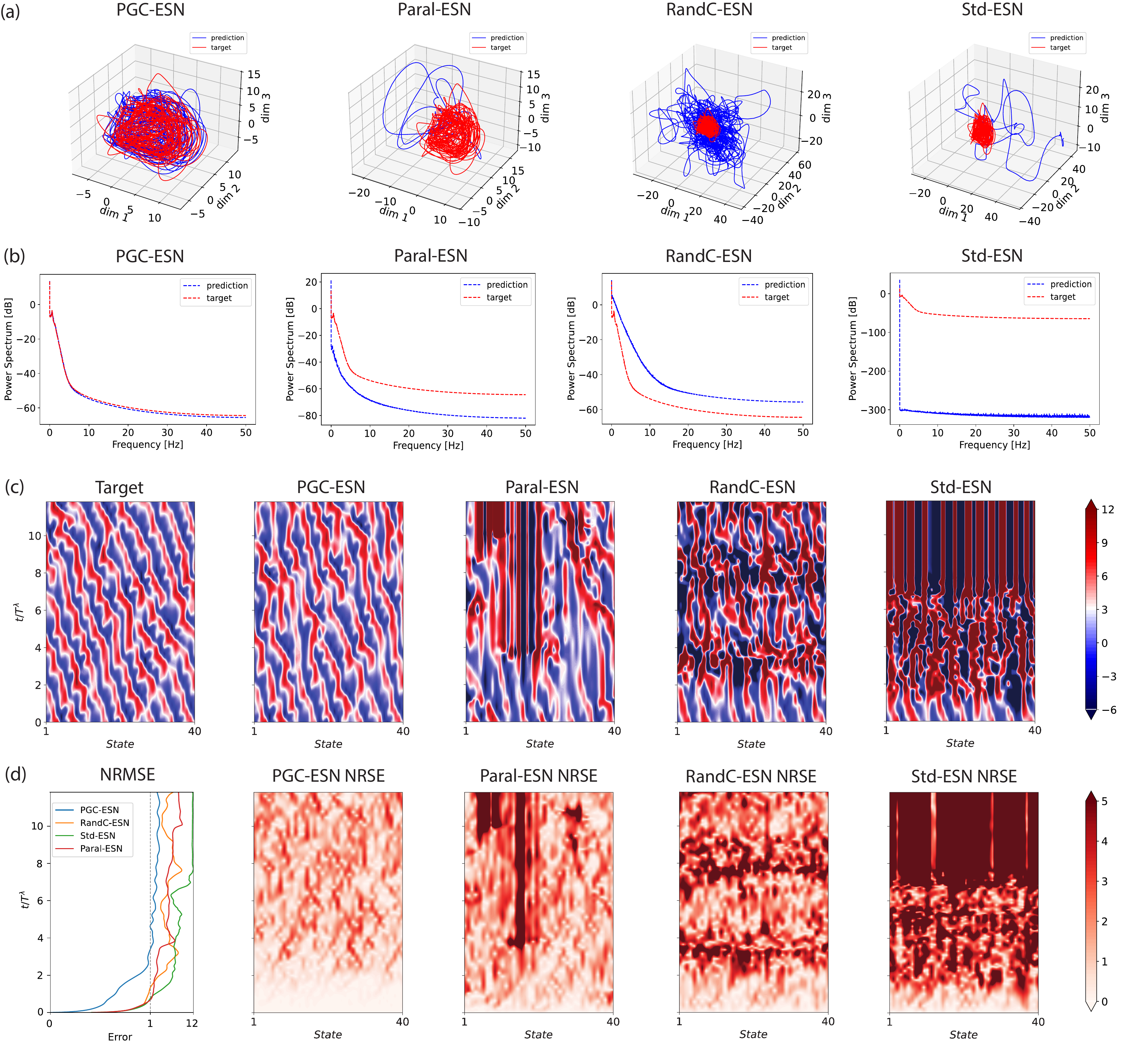}  
    \caption{Predictive result for the Lorenz-96 system using a reservoir size of 500 for PGC-ESN, Paral-ESN, RandC-ESN, and Std-ESN (from left to right). (a) The attractor on the first three-dimensional subspace. (b) The power spectrum. (c) Contour plots of the target and predicted sequences (d) NRMSE evolution and contour plots of the NRSE evolution.
    }
    \label{fig:Lorenz96_reservoir_size_500_plots}
\end{figure*}

Temporal distance~\cite{Hess2023GeneralizedTF} is another invariant statistic, which is used to evaluate the temporal agreement between observed and predicted sequences. It is computed by first measuring the dimension-wise Hellinger distance ($H$) between the power spectra of the two sequences across all state-space dimensions and then averaging the results to obtain a single number, $D_{\mathrm{temp}}$. This metric is suitable for measuring long-term temporal similarity, as it satisfies $0 \leq H \leq 1$, where $H=0$ indicates perfect agreement. Formally, the temporal distance $D_{\mathrm{temp}}$ is defined as
\begin{equation}
  D_{\mathrm{temp}} \coloneqq \frac{1}{d_u}\sum_{i=1}^{d_u} H(f_i, g_i)= \frac{1}{d_u}\sum_{i=1}^{d_u}  \sqrt{1 - \int_{-\infty }^{\infty} \sqrt{f_i(\omega) g_i(\omega)} \, d\omega},
\end{equation}
where $d_u$ is the dimension of the observed system's state space, and $f_i$ and $g_i$ denote the power spectra over frequency $\omega$ for the $i^{\mathrm{th}}$ dimensions of the observed and predicted sequences, respectively. Fig.~\ref{fig:Lorenz96_reservoir_size_500_plots}(b) shows both the target and predicted power spectra in the evaluation of ESN models on the Lorenz-96 system. The $D_{\mathrm{temp}}$ values reported in Table~\ref{tab:Lorenz96_performance} quantify the discrepancy between the power spectra of the target and predicted sequences for each model, averaged across sequences and model initializations.

\subsubsection{\label{subsubsec:Valid_prediction_time}Valid prediction time}
To quantify the short-term predictive accuracy of the trained model over time, we need to compute the normalized root mean square error (NRMSE)~\cite{vlachas2020backpropagation}, which is given by
\begin{equation}
  \text{NRMSE}(\hat{\boldsymbol{u}}(t)) = \sqrt{\left\langle \frac{(\hat{\boldsymbol{u}}(t) - \boldsymbol{u}(t))^2}{\boldsymbol{\sigma}^2(\boldsymbol{u}_{1:T_{\mathrm{ts}}})} \right\rangle},
\end{equation}
where $\boldsymbol{u}(t)$ and $\hat{\boldsymbol{u}}(t)$ are the observed and predicted states at time-step $t$, $\boldsymbol{u}_{1:T_{\mathrm{ts}}}$ is the observed training sequence of length $T_{\mathrm{ts}}$, and $\boldsymbol{\sigma}(\boldsymbol{u}_{1:T_{\mathrm{ts}}})$ denotes the vector of standard deviations over time, computed per state-space dimension of this sequence. The notation $\langle \cdot \rangle$ represents the average over all state-space dimensions. Without this averaging, we obtain the normalized root square error (NRSE), which is a vector spanning all state-space dimensions. These two values will be used to plot the evolution of the model's predictive accuracy.

Another metric used to evaluate predictive accuracy is called the valid prediction time (VPT),~\cite{vlachas2020backpropagation} which is the maximum time $t_m$ for which the model's prediction maintains an NRMSE below a given threshold $\epsilon$, using the Lyapunov time $T^{\lambda}$ as the timescale. Formally, the VPT is given by
\begin{equation}
  \text{VPT} = \max_{t_m} \{ t_m \mid \text{NRMSE}(\hat{\boldsymbol{u}}(t)) < \epsilon, \, \forall t \leq t_m \}.
\end{equation}
In this work, we set the threshold to $\epsilon = 1$ and refer to this metric as VPT-1. This metric is particularly useful for assessing the model's short-term predictive performance.

\subsection{\label{subsec-Experimental_setup}Experimental setup}
Given prior knowledge of the coupling structure in the observed benchmark DS, our goal was to reconstruct the underlying dynamics of the simulated time-series data using four ESN models: PGC-ESN, Std-ESN, RandC-ESN, and Paral-ESN. To evaluate the performance of these ESN models across different reservoir sizes, where the size for Paral-ESN represents the total sum of all reservoirs, we adopted an experimental setup as follows. The observed time series has a length of $T=2 \cdot 10^5 $, which is evenly split into training and test datasets. Within the training set, $40 \% $ is reserved as a validation set during hyperparameter tuning of the input scaling $\beta$ and spectral radius $\rho$, using grid search with $D_{\mathrm{temp}}$ as the selection metric. We add noise to the observed training sequence and standardize it before feeding it into the ESN model during the training phase. The computed mean and standard deviation are also used to scale the observed warm-up sequence during the prediction phase. All evaluation metrics, except for geometric distance, are computed on the descaled observed and predicted sequences. In Std-ESN and RandC-ESN, the densities $p_{\mathrm{in}}$ and $p$ of the input weight matrix $ \mathbf{W}_{in} $ and reservoir weight matrix $ \mathbf{W}$ are set to match those of the PGC-ESN, where both densities are determined by the ratio of coupled dimensions to total dimensions within the observed system. In Paral-ESN, we set $p_{\mathrm{in}}=1$ and $p=1$ to ensure full connectivity within each reservoir unit, aligning with the structure of PGC-ESN. Additionally, the group size $G$ and interaction length $I$ are configured such that each reservoir corresponds to and receives the same number of observed state dimensions as in PGC-ESN, or as close as possible when the coupling structure is asymmetric. We assessed the attractor reconstruction using 5 sequences during tuning and 10 sequences during testing. Each experiment was repeated with 5 randomly initialized models for tuning and 10 for testing, and the final results were averaged. 
All the other hyperparameters are shown in Table~\ref{tab:hyperparameters_for_standard_setting}.

\begin{table}[htbp]
\caption{\label{tab:hyperparameters_for_standard_setting} Hyperparameters for ESNs  }
    \begin{ruledtabular}
        \begin{tabular}{lcr}
            Parameter &Values\\
            \hline
            $T_{\mathrm{warm}}$ & 2000 \\
            $T_{\mathrm{pred}}$ & 7000 \\
            $\rho$ & $\{0.1, 0.4, 0.7, 0.9, 0.99 \}$ \\
            $\beta$ & $\{0.1, 0.5, 1.0, 1.5, 2.0\}$\\
            $\eta$ & 0.01 \\
            $\kappa$ & $5\text{\textperthousand }$
        \end{tabular}
    \end{ruledtabular}
\end{table}

\begin{figure*}[htbp]  
    \centering
    \includegraphics[width=1.0\textwidth]{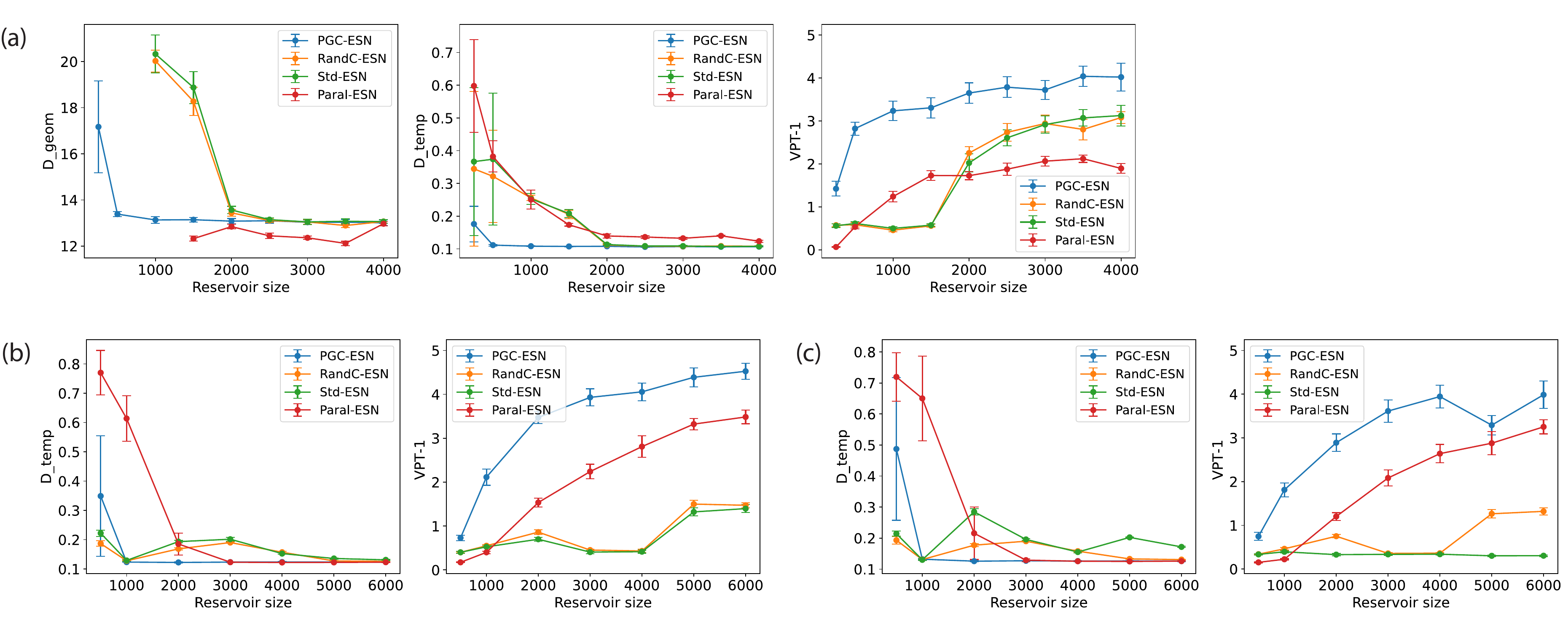}  
    \caption{Model size analysis for the benchmark dynamical systems. (a) The Lorenz-96 system. (b) The homogeneous KS equation. (c) The inhomogeneous KS equation.}
    \label{fig:model_size_analysis_for_all_target_systems}
\end{figure*}

\subsection{\label{Evaluation_results}Evaluation results}

\subsubsection{\label{subsec-Evaluation_results_Lorenz_96}Lorenz-96 system}
The coupling structure of the Lorenz-96 system resulted in a ratio of coupled dimensions to total dimensions of $0.1$. For Paral-ESN, we set $G=1$ and $I=2$. Note that compared to PGC-ESN, which uses the correct coupling structure, each reservoir unit in Paral-ESN receives one additional dimension from the input state on its right-hand side, as illustrated in Fig.~\ref{fig:ODE_and_ESN_models}.
Fig.~\ref{fig:Lorenz96_reservoir_size_500_plots} presents the reconstruction results for a specific target sequence using a reservoir size of 500 for PGC-ESN, Paral-ESN, RandC-ESN, and Std-ESN (from left to right). In Fig.~\ref{fig:Lorenz96_reservoir_size_500_plots}(a), we visualize both the target and predicted attractors on the first three-dimensional subspace, where it is evident that the PGC-ESN achieves the best attractor reconstruction. In Fig.~\ref{fig:Lorenz96_reservoir_size_500_plots}(b), we present the target and predicted power spectra, showing that the PGC-ESN most accurately reproduces the spectral structure, reinforcing its superior performance in capturing long-term dynamics. In Fig.~\ref{fig:Lorenz96_reservoir_size_500_plots}(c), we display contour plots of the first 700 time steps of the target and predicted sequences for all models. Fig.~\ref{fig:Lorenz96_reservoir_size_500_plots}(d) additionally shows the NRMSE evolution averaged over all state-space dimensions and the NRSE evolution across all dimensions. Note that the NRMSE plot uses a logarithmic scale for error on the $x$-axis to emphasize the error range relevant to short-term prediction. The gray horizontal line indicates an error level of 1 and intersects each model’s error trajectory at its corresponding VPT-1 time. The contour plots of the sequences reveal that the PGC-ESN best captures the pattern of the target sequence, while the NRSE plot demonstrates that only the PGC-ESN achieves a clear short-term prediction with a low NRSE error up to approximately 2 Lyapunov times. Moreover, the NRMSE plot shows that PGC-ESN consistently maintains the lowest error throughout the shown time horizon, with a VPT-1 exceeding 2 Lyapunov times, further validating its superior predictive accuracy.
The numerical results for the evaluated metrics, averaged across all sequences and model initializations, are summarized in Table~\ref{tab:Lorenz96_performance}.

\begin{table}[h]
  \centering
  \caption{Comparison of different ESN models's performance with a reservoir size of 500 using the Lorenz-96 system.}
  \begin{ruledtabular}
      \begin{tabular}{l | l| c c c c c}
        System & ESN Model & $D_{\mathrm{geom}}$& $D_{\mathrm{temp}}$ & VPT-1  \\
         \hline
         Lorenz-96
         & PGC-  & \textbf{13.38} $\pm$ \textbf{0.11} &  \textbf{0.11} $\pm$ \textbf{0.003} & \textbf{2.82} $\pm$ \textbf{0.15} \\
         & Std-  & diverging & 0.37 $\pm$ 0.2 & 0.61 $\pm$ 0.05 \\ 
         & Paral-  & diverging & 0.38 $\pm$ 0.05 & 0.53 $\pm$ 0.04 \\
         & RandC-  & diverging & 0.32 $\pm$ 0.14 & 0.58 $\pm$ 0.05 \\
      \end{tabular}
  \end{ruledtabular}
  \label{tab:Lorenz96_performance}
\end{table}

Fig.~\ref{fig:model_size_analysis_for_all_target_systems}(a) presents a comparison of performance across reservoir sizes, starting from 250 and then ranging from 500 to 4000 in increments of 500. Infinite $D_{\mathrm{geom}}$ values resulting from numerical calculations are omitted from the plots. We observe that the PGC-ESN begins achieving a low $D_{\mathrm{geom}}$ value around 13 and $D_{\mathrm{temp}}$ value around 0.1 at a reservoir size of 500, whereas the other three models do not reach comparable performance until a size of 2000. Beyond this point, all models stabilize at similarly low metric values for long-term prediction. This suggests that the PGC-ESN can effectively reconstruct the Lorenz-96 system with a relatively small reservoir, while the other models require significantly larger reservoirs to achieve similar accuracy. The subplot for VPT-1 further highlights the PGC-ESN's superior short-term predictive performance. It achieves clear short-term predictions for reservoir sizes greater than 500, exhibiting noticeably higher VPT-1 values above 3 Lyapunov times. In contrast, Paral-ESN and the other two models only begin demonstrating meaningful short-term predictions at reservoir sizes of 1500 and 2000, respectively, but ultimately plateau at lower values. 
Overall, the proposed PGC-ESN consistently outperforms the other models in long-term attractor reconstruction when the reservoir size is small and in short-term prediction across the entire range of reservoir sizes considered for the Lorenz-96 system. RandC-ESN, which lacks the coupling guidance, performs similarly to Std-ESN, indicating that only introducing a clustered structure in the reservoir layer will not enhance ESN's learning performance, and knowledge of the coupling structure in the target system is the key to PGC-ESN's superior performance.

\begin{figure*}[htbp]  
    \centering
    \includegraphics[width=1.0\textwidth]{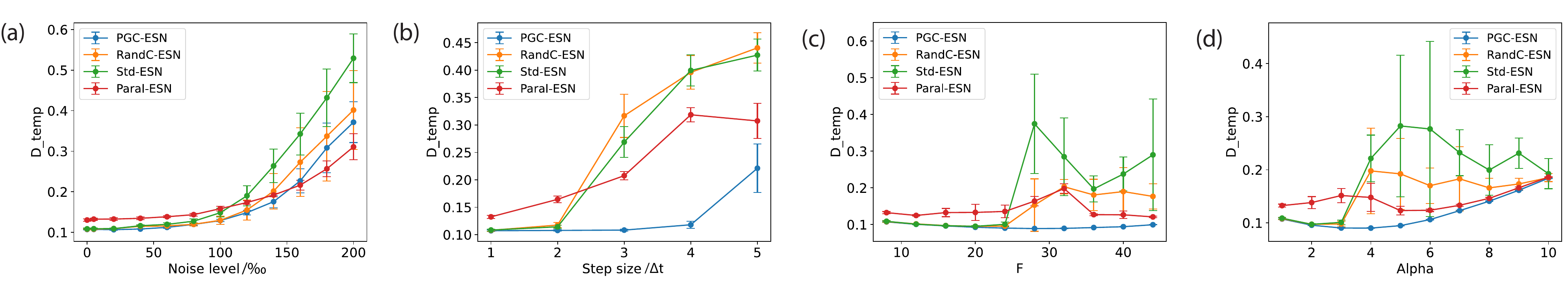}  
    \caption{Evaluation results for the Lorenz-96 system under varied experimental settings. (a) Variation over noise level $\kappa$. 
    (b) Variation over sequence temporal resolution. (c) Variation over the target system's forcing term $F$. (d) Variation over the target system's nonlinearity parameter $\alpha$.
    }
    \label{fig:Lorenz_96_with_varied_experimental_settings}
\end{figure*}

\subsubsection{\label{subsec-Evaluation_results_KS}Kuramoto–Sivashinsky equation}
For both the homogeneous and inhomogeneous Kuramoto–Sivashinsky equations simulated with $d_{u}=256$, we group the dimensions into units of size $G=8$, which is also used as the group size for PGC-ESN and Paral-ESN. Each unit is assumed to be coupled with its adjacent $L=1$ unit, resulting in a ratio of coupled dimensions to total dimensions of approximately 0.1. We set the coupled unit size to $L=1$ for PGC-ESN and the interacting dimension length to $I=8$ for Paral-ESN, ensuring that both models receive the same number of adjacent dimensions as input for each reservoir unit.

The numerical results using a reservoir size of 2000 for all models, averaged across sequences and model initializations, are shown in Table~\ref{tab:KS_performance}, while the corresponding reconstruction plots are provided in Appendix~\ref{Appendix_Kuramoto_Sivashinsky}. These results demonstrate that our proposed PGC-ESN achieves the most accurate long- and short-term predictions for both homogeneous and inhomogeneous KS equations at this reservoir size.

\begin{table}[h]
  \centering
  \caption{Comparison of different ESN models' performance with a reservoir size 2000 using homogeneous (homo) and inhomogeneous (inhomo) Kuramoto–Sivashinsky equations.}
  \begin{ruledtabular}
      \begin{tabular}{l | l| c c c c c}
        System & ESN Model & $D_{\mathrm{temp}}$ & VPT-1  \\
         \hline
         homo KS
         & PGC-    &  \textbf{0.12} $\pm$ \textbf{0.001} & \textbf{3.47} $\pm$ \textbf{0.14} \\
         & Std-    & 0.19 $\pm$ 0.004 & 0.70 $\pm$ 0.04 \\ 
         & Paral-  & 0.19 $\pm$ 0.04 & 1.54 $\pm$ 0.10 \\
         & RandC-  & 0.17 $\pm$ 0.003 & 0.86 $\pm$ 0.05 \\
         
          \hline
         inhomo KS
         & PGC-    &  \textbf{0.13} $\pm$ \textbf{0.001} & \textbf{2.89} $\pm$ \textbf{0.20} \\
         & Std-    & 0.28 $\pm$ 0.01 &  0.33 $\pm$ 0.03 \\ 
         & Paral-  & 0.22 $\pm$ 0.08 & 1.20 $\pm$ 0.09 \\
         & RandC-  & 0.18 $\pm$ 0.005 & 0.75 $\pm$ 0.04 \\
      \end{tabular}
  \end{ruledtabular}
  \label{tab:KS_performance}
\end{table}

Fig.~\ref{fig:model_size_analysis_for_all_target_systems}(b) and (c) present a comparison of performance across reservoir sizes, starting from 500 and then ranging from 1000 to 6000 in increments of 1000 for the homogeneous and inhomogeneous cases, respectively. Note that based on the numerical calculations, all geometric distance values $D_{\mathrm{geom}}$ are infinite and therefore omitted for evaluation on the KS equation. We attribute this to the high dimensionality of the state space for the KS equation, which makes it challenging to effectively approximate the invariant distributions of the observed and reconstructed systems using Gaussian mixture models.

For the homogeneous case shown in Fig.~\ref{fig:model_size_analysis_for_all_target_systems} (b), we observe that PGC-ESN begins to achieve stable and low $D_{\mathrm{temp}}$ values around 0.1 from size 1000, whereas Paral-ESN starts to do so at size 3000, and the other two models exhibit unstable values until size 5000. The VPT-1 plot shows that the PGC-ESN achieves the best short-term predictive power across the whole range, exceeding 4 Lyapunov times for reservoir sizes greater than 3000. In contrast, Paral-ESN and the other two models exceed 3 and 1 Lyapunov times, respectively, for reservoir sizes greater than 4000. A similar trend for the inhomogeneous KS equation is shown in Fig.~\ref{fig:model_size_analysis_for_all_target_systems}(c). Overall, the proposed PGC-ESN consistently outperforms the other models in long-term attractor reconstruction when the reservoir size is small and in short-term prediction across the entire range of reservoir sizes considered for both the homogeneous and inhomogeneous KS equations.

\begin{figure*}[htbp]  
    \centering
    \includegraphics[width=1.0\textwidth]{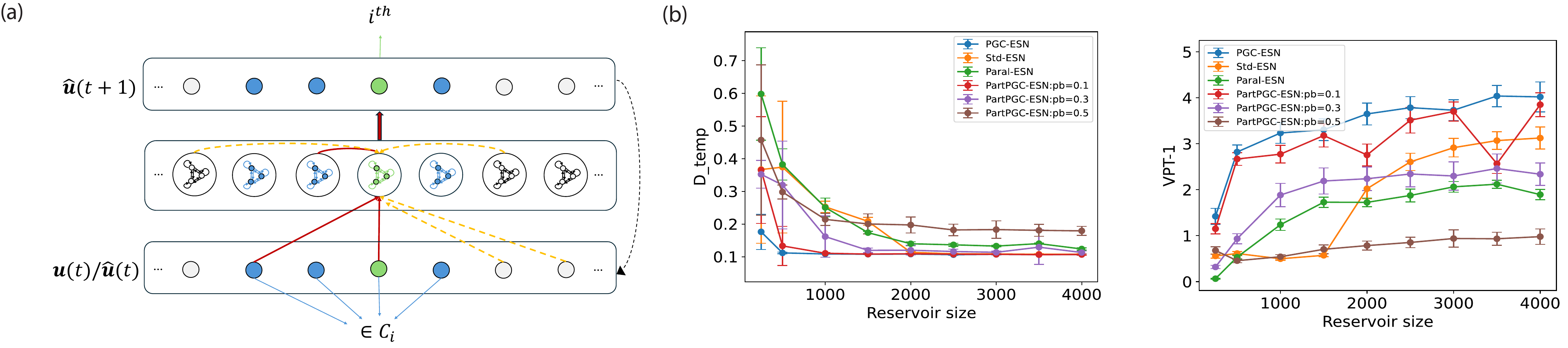}  
    \caption{Investigation of incorporating imperfect prior coupling knowledge. (a) A schematic plot of the PartPGC-ESN with rewired couplings. (b) Evaluation results for the Lorenz-96 system.
    }    \label{fig:Lorenz_96_with_imperfect_prior_coupling_structure_knowledge}
\end{figure*}

\subsection{\label{Lorenz_96_with_varied_experimental_settings}Evaluation on Lorenz-96 system with varied experimental settings}
Based on the evaluation results for learning the Lorenz-96 system across different reservoir sizes, all models reach saturation at a size of 3000. Therefore, we then fixed the reservoir size at 3000 and evaluated the ESN models on the Lorenz96 system under varied experimental settings. The results using the metric $D_{\mathrm{temp}}$ are presented here, while additional metrics can be found in Appendix~\ref{Appendix_Lorenz_96_with_varied_experimental_settings}.

First, we analyzed the robustness of ESN models to noise in training data and to variations in sequence temporal resolution while assessing models tuned under the standard setting. Specifically, we vary the noise levels $\kappa$ from $0$ to $200\text{\textperthousand}$ in increments of $20\text{\textperthousand}$, and simulate varied temporal resolution by downsampling the observed sequences to step sizes ranging from $\Delta t$ to $5\Delta t$ in increments of $\Delta t$, where $\Delta t$ denotes the integration step size used in simulation of the observed trajectory, as described in Subsection~\ref{subsec:Benchmark_Dynamical_Systems}.
Fig.~\ref{fig:Lorenz_96_with_varied_experimental_settings} (a) presents the evaluation results for varying noise level.
We observe that all models' performance deteriorates as the noise level increases, especially beyond 120\text{\textperthousand}. Among them, Paral-ESN and PGC-ESN exhibit a more gradual performance decline, while the other two models deteriorate more noticeably.
Notably, Paral-ESN achieves the lowest $D_{\mathrm{temp}}$ once the noise level exceeds $\kappa = 140\text{\textperthousand}$. We consider that the independent computation of each reservoir's output layer in Paral-ESN may contribute to its enhanced robustness under high noise conditions. Note that PGC-ESN achieves the best short-term prediction performance across the entire range of noise levels, as shown in Appendix~\ref{Appendix_Lorenz_96_with_varied_experimental_settings}.
Overall, the superior robustness of PGC-ESN and Paral-ESN—both of which incorporate prior knowledge of the target system’s coupling structure to some extent—compared to Std-ESN and RandC-ESN, which lack this guidance, suggests that leveraging such physical information plays a key role in mitigating the impact of noise.

Fig.~\ref{fig:Lorenz_96_with_varied_experimental_settings}(b) shows the evaluation results under varying sequence temporal resolutions. We observe that PGC-ESN consistently achieves the best performance across all step sizes, while Paral-ESN exhibits a more gradual performance decline compared to Std-ESN and RandC-ESN, and outperforms both when the step size reaches $3\Delta t$. Additionally, in Appendix~\ref{Appendix_Lorenz_96_with_varied_experimental_settings}, we present the evaluation results for varying training data sizes.

Next, we investigated the impact of systematic nonlinearity by varying the forcing term $F$ and the nonlinearity parameter $\alpha$ in the Lorenz-96 system, tuning and evaluating all models accordingly. 
Specifically, we vary the forcing term $F$ from 8 to 44 in increments of 4 and the nonlinearity parameter $\alpha$ from 1 to 10 in increments of 1. As both $F$ and $\alpha$ increase, the system's maximum Lyapunov exponent (MLE) rises consistently, reaching approximately 9.05 and 14, respectively.
Fig.~\ref{fig:Lorenz_96_with_varied_experimental_settings} (c) presents the evaluation results when the value of $F$ is varied. The results indicate that the PGC-ESN consistently achieves effective attractor reconstruction across the entire range of $F$. In contrast, the other models begin to deteriorate beyond $F = 24$.
Fig.~\ref{fig:Lorenz_96_with_varied_experimental_settings} (d) presents the evaluation results for variation in $\alpha$, also demonstrating that the PGC-ESN consistently outperforms the other models across the entire range of $\alpha$. 
Notably, in both cases, the Paral-ESN exhibits a somewhat more moderate deterioration compared to the Std-ESN and RandC-ESN, which do not incorporate prior coupling knowledge. 

In conclusion, these results further highlight the advantages of incorporating coupling knowledge of the target system into ESN models—such as in our PGC-ESN and Paral-ESN—enhancing their robustness to varying training noise level, training data size, and sequence temporal resolution, while also enabling superior learning performance as the system's nonlinearity increases.

\subsection{\label{Lorenz_96_with_imperfect_prior_coupling_knowledge}Evaluation on Lorenz-96 system with imperfect prior coupling knowledge}

In real-world applications, prior knowledge of the target system's coupling structure is often imperfect. To investigate the impact of such imperfections, we introduced perturbations to the reservoir’s coupling structure of our proposed PGC-ESN model. Specifically, when establishing connections to coupled clusters in the reservoir layer or to coupled input units in the input layer for each reservoir cluster, we randomly rewire each connection to a non-coupled and previously unconnected place, with a given perturbation probability.
 We refer to this perturbed model as the Partially Physics-Guided Clustered ESN (PartPGC-ESN). Fig.~\ref{fig:Lorenz_96_with_imperfect_prior_coupling_structure_knowledge}(a) provides a schematic representation of this model, where the yellow arrows indicate the randomly rewired couplings for the $i^{\mathrm{th}}$ cluster highlighted in green.

We compared the performance of PartPGC-ESN under perturbation probabilities of 0.1, 0.3, and 0.5 with that of the unperturbed PGC-ESN, Std-ESN, and Paral-ESN on the Lorenz-96 system, using the standard settings described in Subsection~\ref{subsec-Experimental_setup}. The evaluation is conducted across various reservoir sizes, starting from 250 and then ranging from 500 to 4000 in increments of 500. Fig.~\ref{fig:Lorenz_96_with_imperfect_prior_coupling_structure_knowledge}(b) presents the plots for $D_{\mathrm{temp}}$ and VPT-1, while the results for $D_{\mathrm{geom}}$ are provided in Appendix~\ref{Appendix_Lorenz_96_with_imperfect_prior_coupling_knowledge}. The results indicate that introducing coupling perturbations degrades the performance of PartPGC-ESN compared to that of the unperturbed PGC-ESN. However, with a perturbation probability of 0.1, PartPGC-ESN still outperforms Std-ESN and Paral-ESN in long-term attractor reconstruction for reservoir sizes smaller than 2000, and in short-term prediction across the entire range of considered reservoir sizes—despite exhibiting slight instability in a few cases. As the perturbation probability increases, the performance of PartPGC-ESN further deteriorates, eventually falling behind that of the Std-ESN and Paral-ESN. These findings suggest that even with imperfect prior coupling knowledge, our proposed approach can still enhance ESN's learning performance in long-term attractor reconstruction under limited model capacity, and in short-term prediction even when the model size is sufficiently large, provided the rate of incorrect couplings remains low.

\begin{figure*}[htbp]  
    \centering
    \includegraphics[width=1.0\textwidth]{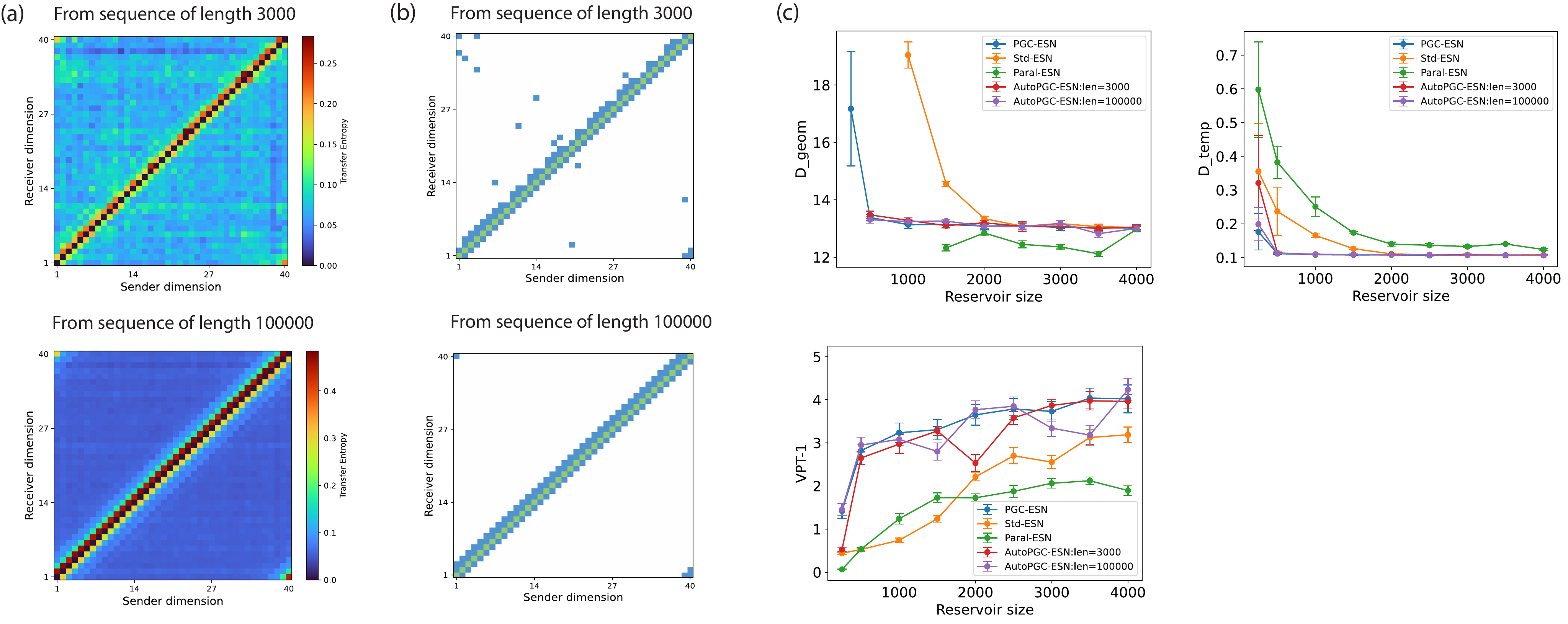}  
    \caption{Investigation of incorporating extracted coupling knowledge. (a) Color map of the coupling strength matrices for the Lorenz-96 system inferred using transfer entropy. (b) Corresponding inferred coupling matrices when using reservoir layer density $p=0.1$.
    (c) Evaluation results for the Lorenz-96 system.
    }
    \label{fig:Lorenz_96_with_no_prior_knowledge}
\end{figure*}

\subsection{\label{Lorenz_96_without_prior_coupling_knowledge}Evaluation on Lorenz-96 system without prior knowledge}

To enhance the practical applicability of our proposed approach, we now consider a scenario in which no prior knowledge of the target system is available. In this setting, the guidance for designing the coupling structure of AutoPGC-ESN’s reservoir is extracted directly from the training data using transfer entropy.
We compare the performance of AutoPGC-ESN with that of Std-ESN on the Lorenz-96 system. In addition to tuning the spectral radius $\rho$ and input scaling $\beta$ over the same range as in the standard setting, we also tune the reservoir layer density $p$ over the set $\{ 0.05,\ 0.1,\ 0.2,\ 0.3 \}$ using grid search, and set the input layer density $p_{\mathrm{in}}$ equal to $p$. This range emphasizes sparse configurations, as sparsity is known to improve the learning performance of ESNs generally.~\cite{lukovsevivcius2012practical}
The other settings follow the standard setup described in Subsection~\ref{subsec-Experimental_setup}.
As the ratio of inferred coupled dimensions (including self-coupling) to total dimensions of the target system is set to $p$, the number of inferred coupled dimensions is then automatically determined by the tuning result.
The evaluation is conducted across various reservoir sizes, starting from 250 and then ranging from 500 to 4000 in increments of 500.
PGC-ESN and Paral-ESN are excluded from this experiment, as they rely on prior knowledge of the target system. For comparison, we also present their performance evaluated under the standard setting, as described in Subsection~\ref{subsec-Evaluation_results_Lorenz_96}.

Fig.~\ref{fig:Lorenz_96_with_no_prior_knowledge}(a) presents heatmaps of the coupling strength matrices inferred for the Lorenz-96 system using transfer entropy, computed on a sequence of length $3 \times 10^3$ and $10^5$ in the training dataset after applying noise addition with noise level $\kappa = 5 \text{\textperthousand }$ and standardization. In each heatmap, the $i^{\mathrm{th}}$ row represents the inferred coupling strengths from all sender dimensions to the $i^{\mathrm{th}}$ receiver dimension, with self-coupling (i.e., the $i^{\mathrm{th}}$ column) set to zero by definition.
When using the sequence of length $3 \times 10^3$, each receiver dimension receives the strongest inferred coupling from its immediate neighbors, while the coupling strengths from non-adjacent dimensions are weaker and exhibit slight variation in magnitude.
In contrast, when using the sequence of length $10^5$, the heatmap shows a clearer structure: each receiver dimension receives the dominant coupling strengths from the true coupled dimensions, followed by weaker but similarly valued coupling strengths from their neighboring dimensions. Coupling strengths from all other dimensions are weak and relatively uniform in magnitude.

Fig.~\ref{fig:Lorenz_96_with_no_prior_knowledge}(b) shows the inferred coupling matrices, where for each receiver dimension (highlighted in green), the inferred coupled dimensions are indicated in blue. The results are obtained using a reservoir layer density of $p = 0.1$, which yields three (non-self) inferred coupled dimensions per row, matching the true number of non-self couplings of the target system.
Using time series of lengths $3 \times 10^3$ and $10^5$ for coupling extraction, the overall accuracies of correctly identified non-self couplings relative to the ground-truth coupling structure reach $89.17\%$ and $100\%$, respectively.
These results indicate that longer training sequences facilitate more accurate inference of the underlying coupling structure, capturing a richer representation of the target system’s intrinsic dependencies.

Fig.~\ref{fig:Lorenz_96_with_no_prior_knowledge} (c) presents the evaluation results.
AutoPGC-ESN is evaluated using coupling knowledge extracted from training sequences of length $3 \times 10^3$ and $10^5$. When the reservoir size is as small as 250, AutoPGC-ESN, leveraging a longer training sequence for coupling extraction, achieves performance comparable to PGC-ESN and outperforms the variant using a shorter sequence in both $D_{\mathrm{temp}}$ and VPT-1.
Using either near-perfect coupling knowledge extracted from sequences of length $10^5$ or even imperfect coupling knowledge extracted from a sequence of length $3 \times 10^3$, AutoPGC-ESN consistently outperforms Std-ESN and Paral-ESN in long-term attractor reconstruction for reservoir sizes below 1500, as well as in short-term prediction across the entire range of reservoir sizes considered, despite exhibiting minor instability in a few instances.
These results indicate that the coupling knowledge automatically extracted and utilized by our proposed approach, whether near-perfect or slightly imperfect, can still significantly enhance the ESN’s learning performance in long-term attractor reconstruction under limited model capacity, and in short-term prediction even when the model size is sufficiently large. 
Overall, these findings highlight the practical utility of our method even when prior knowledge of the target system is unavailable.

\section{\label{sec:conclusion}Conclusion}
Our work demonstrates that the coupling knowledge from an observed dynamical system can serve as a powerful guide to designing ESNs with a similarly coupled clustered reservoir layer. By conducting evaluations using benchmark dynamical systems, including the Lorenz-96 system and both homogeneous and inhomogeneous Kuramoto-Sivashinsky equations, we validate the effectiveness of our proposed PGC-ESN for both long-term attractor reconstruction and short-term prediction, even when the prior or extracted coupling knowledge is slightly imperfect. More specifically, incorporating coupling knowledge into ESNs leads to a substantial reduction in computational resources, such as reservoir size and the amount of training data. When resources are abundant, alternative approaches eventually match PGC-ESN in long-term attractor reconstruction but continue to underperform in short-term prediction.
We attribute this trend to the inherent chaoticity of the underlying system, where sensitivity to initial conditions renders short-term prediction particularly challenging. In this context, incorporating prior coupling knowledge enables PGC-ESN to achieve superior state forecasting, even when computational resources are abundant. By contrast, the metrics used for long-term attractor reconstruction quantify the geometric and temporal discrepancies of the target and predicted sequences averaged over a prolonged time horizon. As a result, PGC-ESN’s advantage in this aspect diminishes once performance saturation is reached across all ESN models under sufficient computational resources.

By comparing PGC-ESN against a model with a clustered reservoir that lacks a coupling structure (RandC-ESN), we demonstrate that the effectiveness of PGC-ESN arises from the coupling structure in the reservoir. Additionally, by evaluating the models on the Lorenz-96 system with varied experimental settings, we show that compared to Std-ESN and RandC-ESN—which do not utilize coupling knowledge—incorporating coupling knowledge into ESN models, as done in PGC-ESN and Paral-ESN, enhances their robustness to variations in training noise level, training data size, and sequence temporal resolution, while also enabling superior predictive performance as the target system's nonlinearity increases.

We argue that the target system's coupling knowledge serves as an inductive bias in the learning process, effectively regularizing the model and constraining the hypothesis space to a more optimal region for learning DSs. A theoretical analysis of this phenomenon remains an open direction for future research. Moreover, using a more complex coupling structure in structured data as an inductive bias presents another exciting research direction. For example, in a link prediction task for a system with a network structure, like the social network and chemical reaction network, the system’s coupling knowledge could be utilized to guide the design of the ESN’s reservoir structure and other ML models. 

However, our method has a restriction on computational efficiency compared to Paral-ESN. 
Paral-ESN is known for its superior scalability, owing to its modular architecture and the possibility of sequential training, which makes it particularly well-suited for learning large-scale spatiotemporal systems, such as 2- or 3-dimensional PDE systems. For homogeneous systems, training a single reservoir and replicating it spatially is often sufficient when using Paral-ESN.~\cite{vlachas2020backpropagation} Given the computational efficiency of Paral-ESN and prior work incorporating fully coupled structures into Paral-ESN for learning chaotic systems,~\cite{baur2021predicting} we believe that integrating the system's coupling knowledge into Paral-ESN's framework represents a promising direction for future research.
Moreover, by leveraging spatial translation symmetry, a trained Paral-ESN can generalize to systems with varying spatial extent.~\cite{goldmann2022learn} Prior studies have also demonstrated Paral-ESN’s flexibility to integrate with dimensionality reduction techniques.~\cite{fleddermann2025improving} These insights suggest that incorporating system-specific knowledge, such as coupling structure, into ESN models in conjunction with the aforementioned and more general techniques for ESNs holds significant potential for advancing future research.

Our results also suggest a future direction in which the reservoir could self-organize to optimize its network configuration. In this study, we showed that even without prior knowledge of the coupling structure, incorporating coupling information extracted via external time-series analysis enhances ESN's predictive performance. Banerjee et al.~\cite{banerjee2021machine} demonstrated that a reservoir model can perform link inference for dynamics directly from time series data. This implies that the reservoir itself could simultaneously extract coupling knowledge from time series data and use it to self-organize its internal structure accordingly. Furthermore, the observed contribution of the similarity between the reservoir and target system structures for predictive performance may help explain why physical reservoir computing~\cite{nakajima2020physical} excels at predicting and controlling the dynamics of the physical system itself.~\cite{goto2021twin, akashi2024embedding}

Beyond this study, we hope our work will inspire future research on integrating coupling knowledge and broader physical principles into the architectures of artificial neural networks and physical reservoir computers. Physical systems are subject to various constraints, such as spatial limitations and input-channel restrictions, which affect the coupling between computational nodes. These constraints make it challenging to configure full connectivity and modulate existing connections. Therefore, designing computational architectures tailored to specific tasks is essential for physical computational systems. The findings of this study provide valuable insights into the relationship between information processing and computational medium.

\begin{acknowledgments}
N. A. was supported by JSPS KAKENHI Grant No. JP22KJ1786, 25K00011, and JST PREST Grant No. JPMJPR24KF.
\end{acknowledgments}

\section*{Conflict of Interest}
The authors have no conflicts to disclose.

\section*{Data Availability}
The data that support the findings of this study are available at \url{https://github.com/kuei-jan-chu/physics_guided_clustered_ESN}.

\appendix

\section{\label{sec:Additional_figures_of_ESN_models}Additional figures of ESN models}

\subsection{\label{Appendix_Learning_process_for_ESNs}Learning process for ESNs}
\begin{figure}[htbp]  
    \centering
\includegraphics[width=0.45\textwidth]{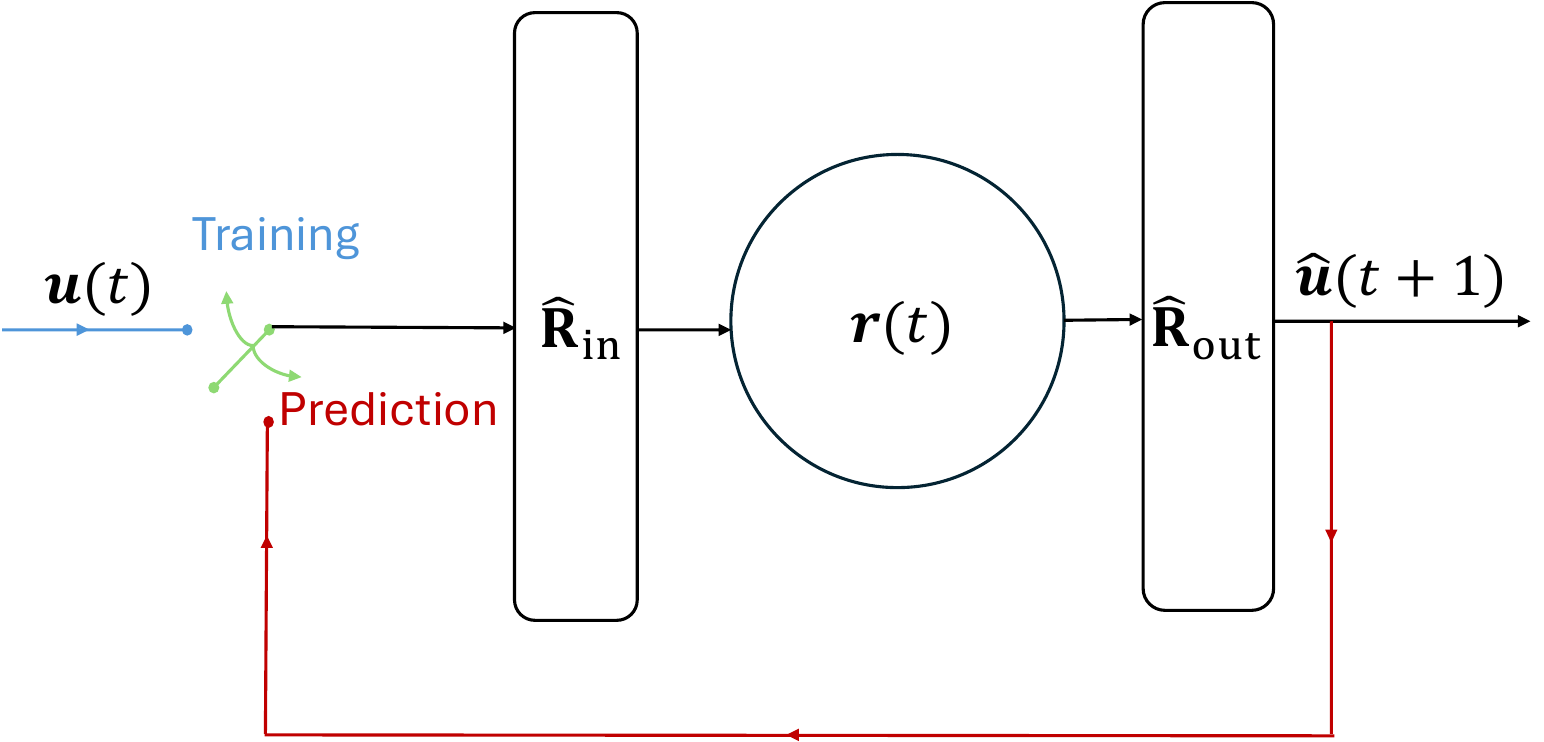}  
    \caption{Schematic diagram of the ESN learning process
    }
    \label{fig:ESN_learning_process}  
\end{figure}

We present a schematic diagram of the ESN learning process in Fig.~\ref{fig:ESN_learning_process}. The ESN consists of an input layer $\hat{\mathbf{R}}_\mathrm{in}$, a reservoir layer with state $\boldsymbol{r}$, and an output layer $\hat{\mathbf{R}}_\mathrm{out}$. During the training phase, the reservoir receives the observed system state  $\boldsymbol{u}(t)$ as input. After sufficient training, the output layer maps the reservoir state $\boldsymbol{r}(t)$ to the predicted state $\hat{\boldsymbol{u}}(t+1)$ for the next time step. In the prediction phase, the model operates autonomously, using its own predicted states as input to generate sequences, thereby reconstructing the observed dynamical system (DS).

\subsection{\label{sec:appendix_PGC-ESN_for_PDE_system}Physics-guided clustered ESNs for PDE target systems}
\begin{figure}[htbp]  
    \centering
    \includegraphics[width=0.5\textwidth]{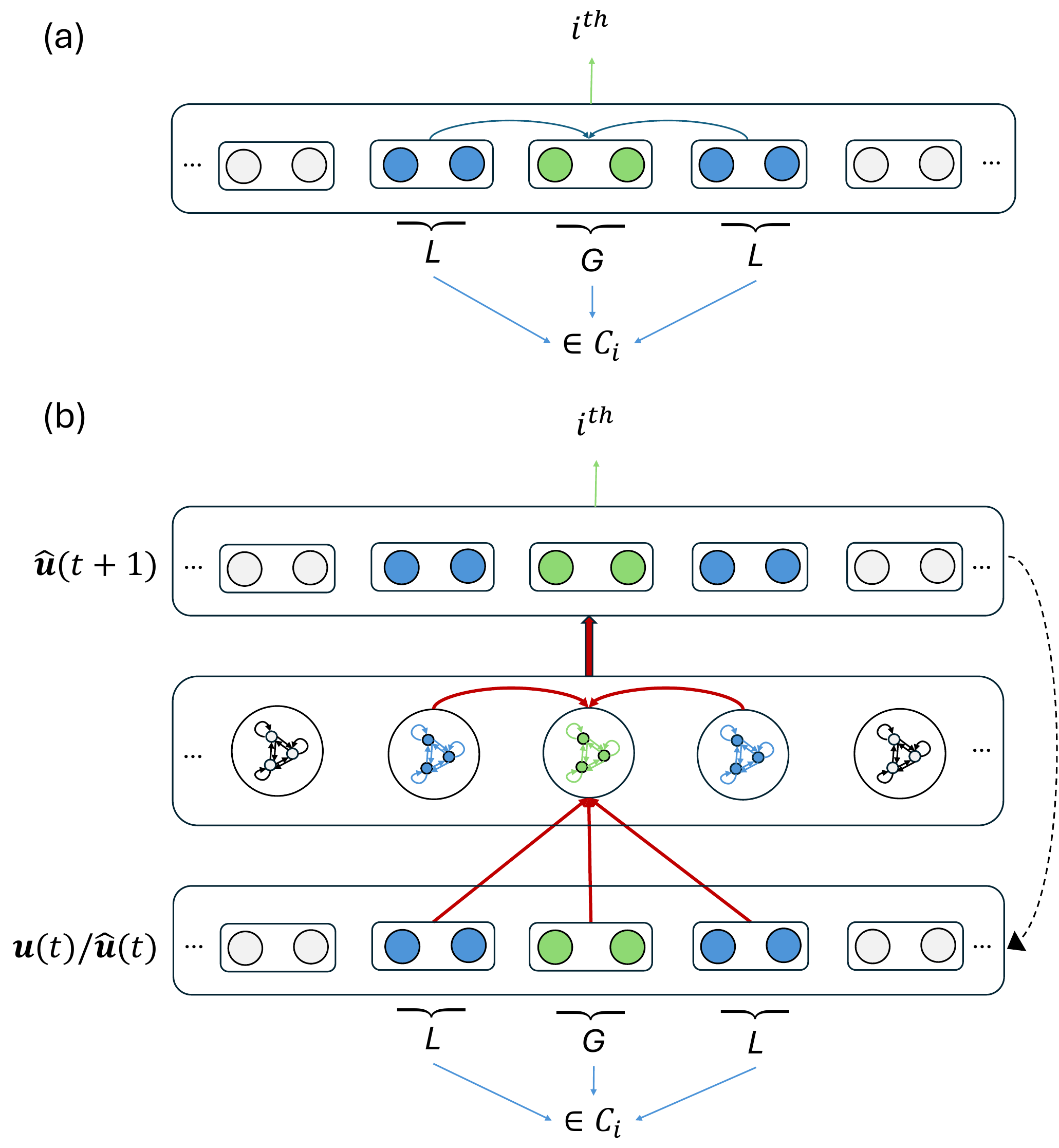}  
    \caption{Illustrative diagram of an example target PDE system (a) and the learning process using the corresponding PGC-ESN (b).}
    \label{fig:PDE_and_PGClustered_ESN}  
\end{figure}

Fig.~\ref{fig:PDE_and_PGClustered_ESN}(a) shows an example of a PDE system where we group the dimensions into units with size $G=2$. The specific $i^{\mathrm{th}}$ unit highlighted in green is symmetrically coupled with the $L=1$ neighboring unit colored in blue on both the left and right sides. Therefore, the index set for the coupled units of the $i^{\mathrm{th}}$ unit here is given by $C_i = \{ i-1, i, i+1 \}$. The coupling structure for all other units is identical and omitted here. Fig.~\ref{fig:PDE_and_PGClustered_ESN} (b) illustrates the corresponding PGC-ESN architecture for this PDE system. For clarity, we highlight a specific unit in the input and predicted states and their associated cluster in the PGC-ESN in green, while the coupled units and clusters are shown in blue.

\subsection{\label{Appendix_Reservoir_updation_for_PGC_ESNs}Reservoir updation for physics-guided clustered ESNs}
\begin{figure}[htbp]  
    \centering
    \includegraphics[width=0.5\textwidth]{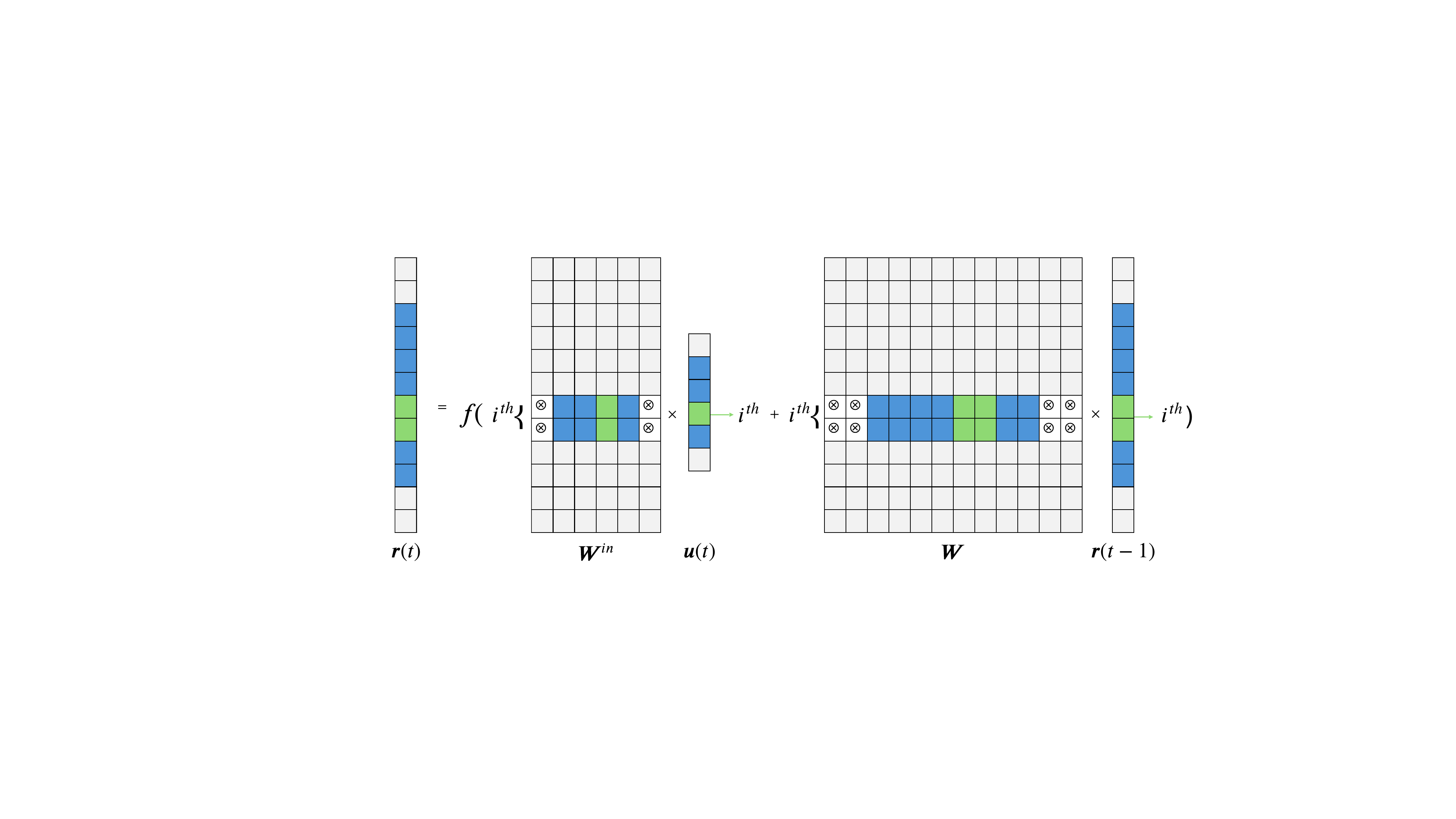}  
    \caption{Updation rule for the PGC-ESN}
    \label{fig:updation_for_pgclustered_esn}  
\end{figure}

Fig.~\ref{fig:updation_for_pgclustered_esn} provides a schematic illustration of the reservoir state update rule for PGC-ESNs with a reservoir state $\boldsymbol{r} \in \mathbb{R}^{12}$ when applied to a dynamical system with state $\boldsymbol{u} \in \mathbb{R}^6$. We focus on the updation of the $i^{\mathrm{th}}$ cluster in the reservoir state. The $i^{\mathrm{th}}$ dimension of the input state, along with the corresponding cluster in the reservoir, are highlighted in green, while the coupled dimensions and clusters are shown in blue. In both matrices $\boldsymbol{W}^{\mathrm{in}}$ (the input layer) and $\boldsymbol{W}$ (the reservoir layer), the symbol $\otimes$ indicates entries with a value of 0, and the blue grids represent the connections that transmit information from the coupled dimensions in $\boldsymbol{u}(t)$  or the coupled clusters in $\boldsymbol{r} (t-1)$. Meanwhile, the green grids transmit information from the corresponding dimension in $\boldsymbol{u}(t)$ or the cluster itself in $\boldsymbol{r}(t-1)$.

\section{\label{sec:appendix_transfer_entropy_details}Definition and numerical details on transfer entropy}

Transfer Entropy (TE) is an information-theoretic measure that quantifies the directed information flow between two time series.~\cite{schreiber2000measuring} Let $X_t$ and $Y_t$ for $t \in \mathbb{N}$ be two stochastic processes, the transfer entropy from $Y$ to $X$ is defined as the conditional mutual information between the next state $X_{t+1}$ and the past of $Y$, conditioned on the past of $X$:
\begin{align}
T_{Y \to X} &= I(X_{t+1}; \boldsymbol{Y}_t^{(L)} \mid \boldsymbol{X}_t^{(L)}) \nonumber \\ 
&= \iiint p(x_{t+1}, \boldsymbol{x}_t^{(L)}, \boldsymbol{y}_t^{(L)}) \log \left( \frac{p(x_{t+1} \mid \boldsymbol{x}_t^{(L)}, \boldsymbol{y}_t^{(L)})}{p(x_{t+1} \mid \boldsymbol{x}_t^{(L)})} \right) \, \nonumber \\ 
&\quad \quad \quad dx_{t+1} \, d\boldsymbol{x}_t^{(L)} \, d\boldsymbol{y}_t^{(L)}
\end{align}
where $I(\cdot\,;\,\cdot \mid \cdot)$ denotes the conditional mutual information, $\boldsymbol{X}_t^{(L)} = [X_{t-L+1}, \dots, X_{t-1}, X_t]$ is the length-$L$ past state vector of $X$, and $ \boldsymbol{Y}_t^{(L)} = [Y_{t-L+1}, \dots, Y_{t-1}, Y_t]$ is the length-$L$ past state vector of $Y$.

In our implementation, the calculation of transfer entropy is restricted to pairs of univariate time series, as it is used solely to infer the coupling strengths between individual dimensions of the Lorenz-96 system in this study. We set the length of the past state vectors to $L=5$ and retain only the earliest state in each vector. To approximate the underlying probability distributions, each time series is discretized using histogram binning, with the number of bins automatically determined via the Freedman–Diaconis rule.~\cite{freedman1981histogram} To reduce estimation noise, the approximated probability distributions are smoothed using a Gaussian filter with standard deviation $\sigma = 1$. The final integration is approximated by summing the smoothed probabilities over all bins.

\section{\label{sec:appendix_evaluation_metrics_details}Numerical details on the evaluation metrics}
\subsection{\label{subsec:appendix_geometrical_distance_details}Geometrical distance}

To ensure that the measure is evaluated according to the system's attractor rather than transient dynamics, we discard the first $25 \%$ of time steps in both the observed and predicted sequences to obtain sequences of length $T$ for evaluation.
The ideal invariant distributions $p(\boldsymbol{u})$ and $p(\boldsymbol{u} | \boldsymbol{\theta})$, which represent the state space densities of the ground truth and reconstructed systems, are approximated via Gaussian mixture models (GMMs) using the observed and predicted sequences $\{ \boldsymbol{u}(1), \dots \boldsymbol{u}(T) \}$ and $\{ \hat{\boldsymbol{u}}(1), \dots \hat{\boldsymbol{u}}(T) \}$, respectively:
\begin{equation}
    \begin{aligned}
        p(\boldsymbol{u}) &\approx \hat{p}(\boldsymbol{u}) = \frac{1}{T} \sum_{t=1}^T \mathcal{N}(\boldsymbol{u} ; \boldsymbol{u}(t), \boldsymbol{\Sigma}), \\
        p(\boldsymbol{u} | \boldsymbol{\theta} ) &\approx \hat{p}(\boldsymbol{u} | \boldsymbol{\theta} ) = \frac{1}{T} \sum_{t=1}^T \mathcal{N}(\boldsymbol{u} ; \hat{\boldsymbol{u}}(t), \boldsymbol{\Sigma}),
    \end{aligned}
\end{equation}
where each Gaussian distribution is centered at an observed or predicted state, respectively; the covariance matrix $\boldsymbol{\Sigma}$ determines the granularity of the spatial resolution. We set $\boldsymbol{\Sigma} = \mathbf{I}$ here, following the setting in Ref.~\onlinecite{Hess2023GeneralizedTF}. The Kullback-Leibler (KL) divergence is then estimated using a Monte Carlo approximation:
\begin{equation}
  D_{\mathrm{geom}} \coloneqq  D_{\mathrm{KL}} (p(\boldsymbol{u}) \Vert p(\boldsymbol{u} | \boldsymbol{\theta}))
  \approx \frac{1}{n} \sum_{i=1}^n \log \frac{\hat{p}(\boldsymbol{u}(i))}{\hat{p}(\boldsymbol{u}(i) | \boldsymbol{\theta} )},
\end{equation}
where $n=1000$ Monte Carlo samples $\boldsymbol{u}(i)$ are drawn individually from a randomly selected individual Gaussian component of the GMM approximating $\hat{p}(\boldsymbol{u})$.

\begin{figure*}[htbp]  
    \centering
    \includegraphics[width=1.0\textwidth]{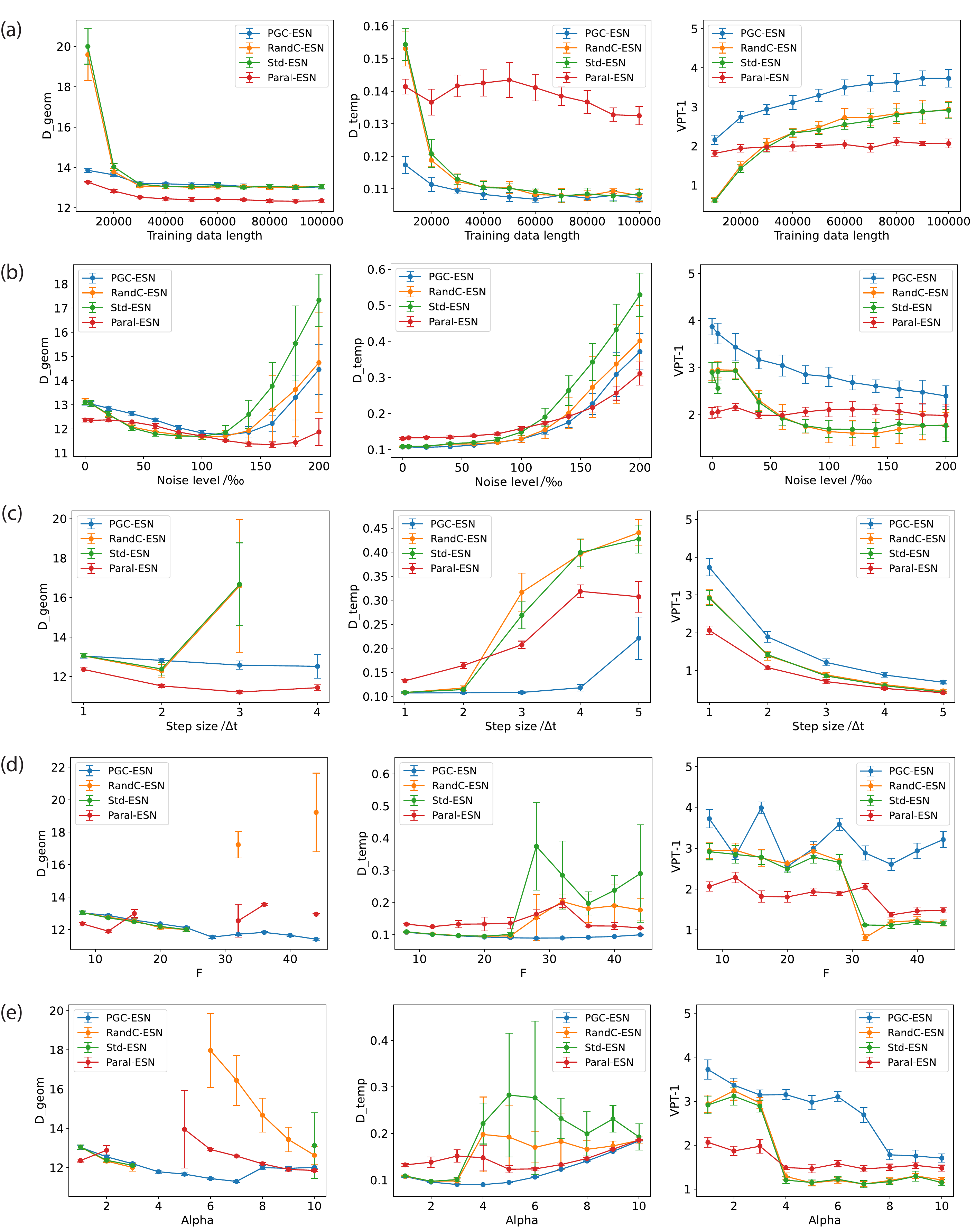}  
    \caption{Evaluation results for the Lorenz-96 system under varied experimental settings. (a) Variation over training data size. (b) Variation over noise level $\kappa$. (c) Variation over sequence temporal resolution.
    (d) Variation over the target system’s forcing term  $F$. (e) Variation over the target system’s nonlinearity parameter $\alpha$.
    }
    \label{fig:all_Lorenz_96_with_varied_experimental_settings}
\end{figure*}

\subsection{\label{subsec:appendix_temporal_distance_details}Temporal distance} 

As with the geometrical distance $D_{\mathrm{geom}}$, we discard the first $25 \%$ of time steps in the observed and predicted sequences to eliminate the influence of transient dynamics. We then standardize each dimension of both sequences $\{ \boldsymbol{u}(1), \dots \boldsymbol{u}(T) \}$ and $\{ \hat{\boldsymbol{u}}(1), \dots \hat{\boldsymbol{u}}(T) \}$ so that $f_i$ and $g_i$ satisfy $\int_{-\infty }^{\infty} f_i(\omega) \, d\omega = 1$ and $\int_{-\infty }^{\infty} g_i(\omega) \, d\omega = 1$, which is required to compute the Hellinger distance $H$. We approximate the power spectra $f_i$ and $g_i$ using a Fast Fourier Transform (FFT)~\cite{cooley1965algorithm}, yielding $\hat{\boldsymbol{f}}_i = |\mathcal{F} u_{i, 1:T}|^2$ and $\hat{\boldsymbol{g}}_i = |\mathcal{F} \hat{u}_{i, 1:T}|^2$, where $\hat{\boldsymbol{f}}_i$ and $\hat{\boldsymbol{g}}_i$ are the discrete power spectra for the $i^{\mathrm{th}}$ dimension. The approximated power spectra are then smoothed using a Gaussian kernel with a standard deviation $\sigma = 2$ and a window length $l = 10\sigma + 1$. The high-frequency tails beyond the first 500 frequency components are cut off. Finally, the trapezoidal rule is applied to numerically integrate the pointwise geometric mean of the discrete power spectra over frequency.

\begin{figure*}[htbp]  
    \centering
    \includegraphics[width=1.0\textwidth]{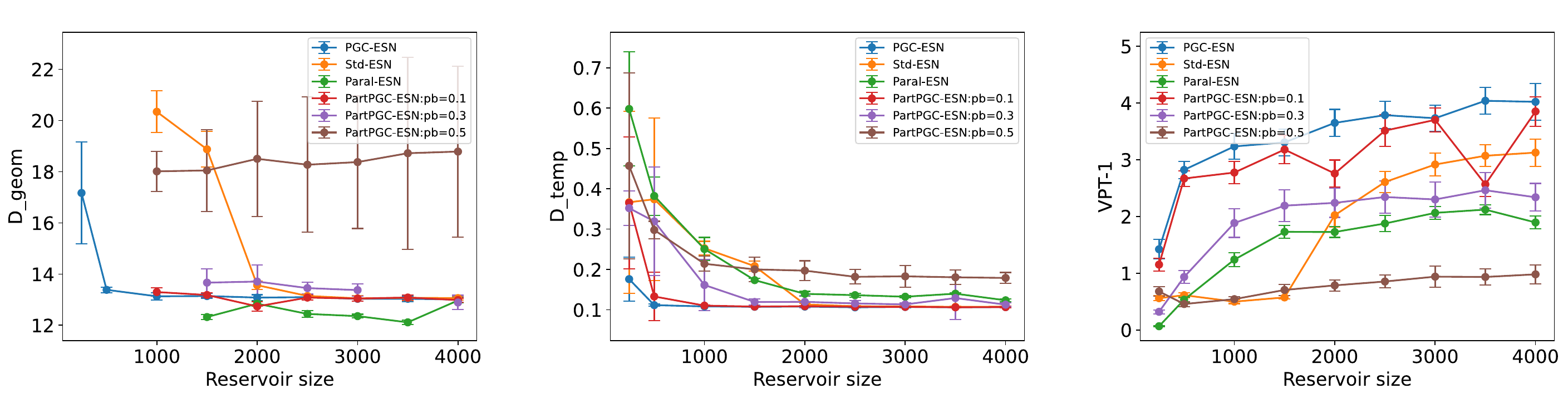}  
    \caption{Evaluation results for incorporating imperfect prior coupling knowledge using the Lorenz-96 system.}
    \label{fig:all_results_for_Lorenz_96_with_imperfect_prior_coupling_structure_knowledge}
\end{figure*}

\section{\label{Appendix_More_evaluation_results}More evaluation results}
\subsection{\label{Appendix_Lorenz_96_with_varied_experimental_settings}Lorenz-96 with varied experimental settings}
Here, we present all evaluation results for the models under varied experimental settings with a reservoir size of 3000. First, we evaluate models tuned under standard settings but trained with different data sizes, noise levels $\kappa$, and sequence temporal resolution.

Fig.~\ref{fig:all_Lorenz_96_with_varied_experimental_settings} (a) presents the evaluation results across different training data sizes, ranging from $10^5$ down to $10^4$ in decrements of $10^4$. The findings reveal that our proposed PGC-ESN effectively reconstructs the system dynamics, maintaining $D_{\mathrm{geom}}$ around 14 and $D_{\mathrm{temp}}$ around 0.12, while achieving short-term prediction with VPT-1 around 2 Lyapunov times, even with a reduced training data size of $10^4$. Paral-ESN also exhibits moderate sensitivity to reductions in training data size, achieving the lowest $D_{\mathrm{geom}}$ values across the entire range, while maintaining inferior performance in $D_{\mathrm{temp}}$ and VPT-1 with mild deterioration along the data reduction. In contrast, the predictive accuracy of the other two models without incorporating coupling knowledge deteriorates abruptly when the training data size falls below $3 \times 10^4$. These results demonstrate that incorporating the coupling structure of the target system—as done in PGC-ESN and Paral-ESN—can enhance the ESN's learning ability when the training data size is small.

Fig.~\ref{fig:all_Lorenz_96_with_varied_experimental_settings} (b) presents the evaluation results under varying noise levels added to the training dataset. Similar to the $D_{\mathrm{temp}}$ plot described in Sec.~{IV E} of the main paper, we can also see from the $D_{\mathrm{geom}}$ and VPT-1 plots that the performance of Paral-ESN and PGC-ESN declines more gradually than the other two models without incorporating the target system's coupling structure. Additionally, our proposed PGC-ESN achieves the best short-term predictive power across the whole range of noise levels.

Fig.~\ref{fig:all_Lorenz_96_with_varied_experimental_settings} (c) presents the evaluation results under varying sequence temporal resolutions. Consistent with the $D_{\mathrm{temp}}$ plot discussed in Sec.~{IV E} of the main paper, the $D_{\mathrm{geom}}$ results indicate that PGC-ESN and Paral-ESN exhibit a more gradual decline in long-term attractor reconstruction performance as the step size increased, compared to Std-ESN and RandC-ESN. In contrast, short-term prediction performance deteriorates at a similar rate across all models. Note that when the step size reaches $5\Delta t$, the $D_{\mathrm{geom}}$ values for all models diverge to infinity and are therefore omitted from the plot.

Next, we tune and evaluate all models on the Lorenz-96 system under varying values of the forcing term $F$ and the nonlinearity parameter $\alpha$. Fig.~\ref{fig:all_Lorenz_96_with_varied_experimental_settings}(d) and (e) present all the evaluation results for variations in $F$ and $\alpha$ of the target system, respectively. In (d), we observe that similar to the $D_{\mathrm{temp}}$ plots described in Sec.~{IV E} of the main paper, the $D_{\mathrm{geom}}$ and VPT-1 plots show that our proposed PGC-ESN also achieves effective long-term attractor reconstruction and short-term prediction across the entire range of $F$, while the other models deteriorate as $F$ increases. 
In (e), it is evident that our proposed PGC-ESN excels in all three metric plots. Additionally, from the VPT-1 plot, we can see that PGC-ESN achieves effective short-term prediction up to $\alpha = 6$, with over 3 Lyapunov times, while the other models deteriorate noticeably after $\alpha = 3$.

\subsection{\label{Appendix_Lorenz_96_with_imperfect_prior_coupling_knowledge}Lorenz-96 with imperfect prior coupling knowledge}

Fig.~\ref{fig:all_results_for_Lorenz_96_with_imperfect_prior_coupling_structure_knowledge} presents all the evaluation results for investigating the impact of using imperfect prior coupling knowledge in the Lorenz-96 system to guide the PGC-ESN's reservoir structure. The plots for $D_{\mathrm{temp}}$ and VPT-1 are illustrated in Sec.~{IV F} of the main paper. Additionally, the $D_{\mathrm{geom}}$ plot shows the same trend that we discussed there.

\subsection{\label{Appendix_Kuramoto_Sivashinsky}Kuramoto–Sivashinsky} 
\begin{figure*}[htbp]  
    \centering
    \includegraphics[width=1\textwidth]{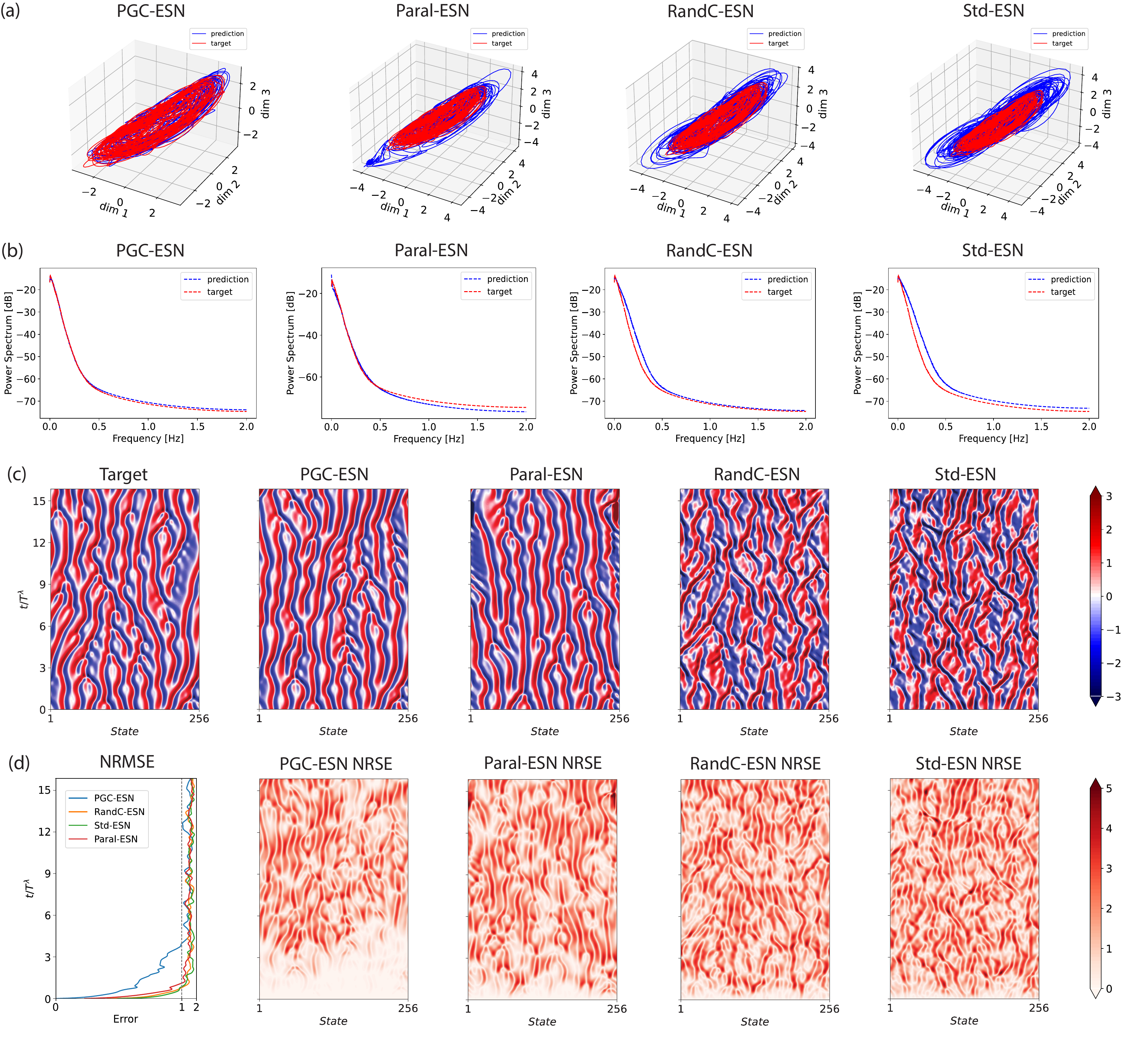}  
    \caption{Predictive results for the homogeneous KS equation using a reservoir size of 2000 for PGC-ESN, Paral-ESN, RandC-ESN, and Std-ESN (from left to right). (a) The attractor on the first three-dimensional subspace. (b) The power spectrum. (c) Contour plots of the target and predicted sequences (d) NRMSE evolution and contour plots of the NRSE evolution.}
    \label{fig:homo_KS_reservoir_size_2000_plots}
\end{figure*}

\begin{figure*}[htbp]  
    \centering
    \includegraphics[width=1\textwidth]{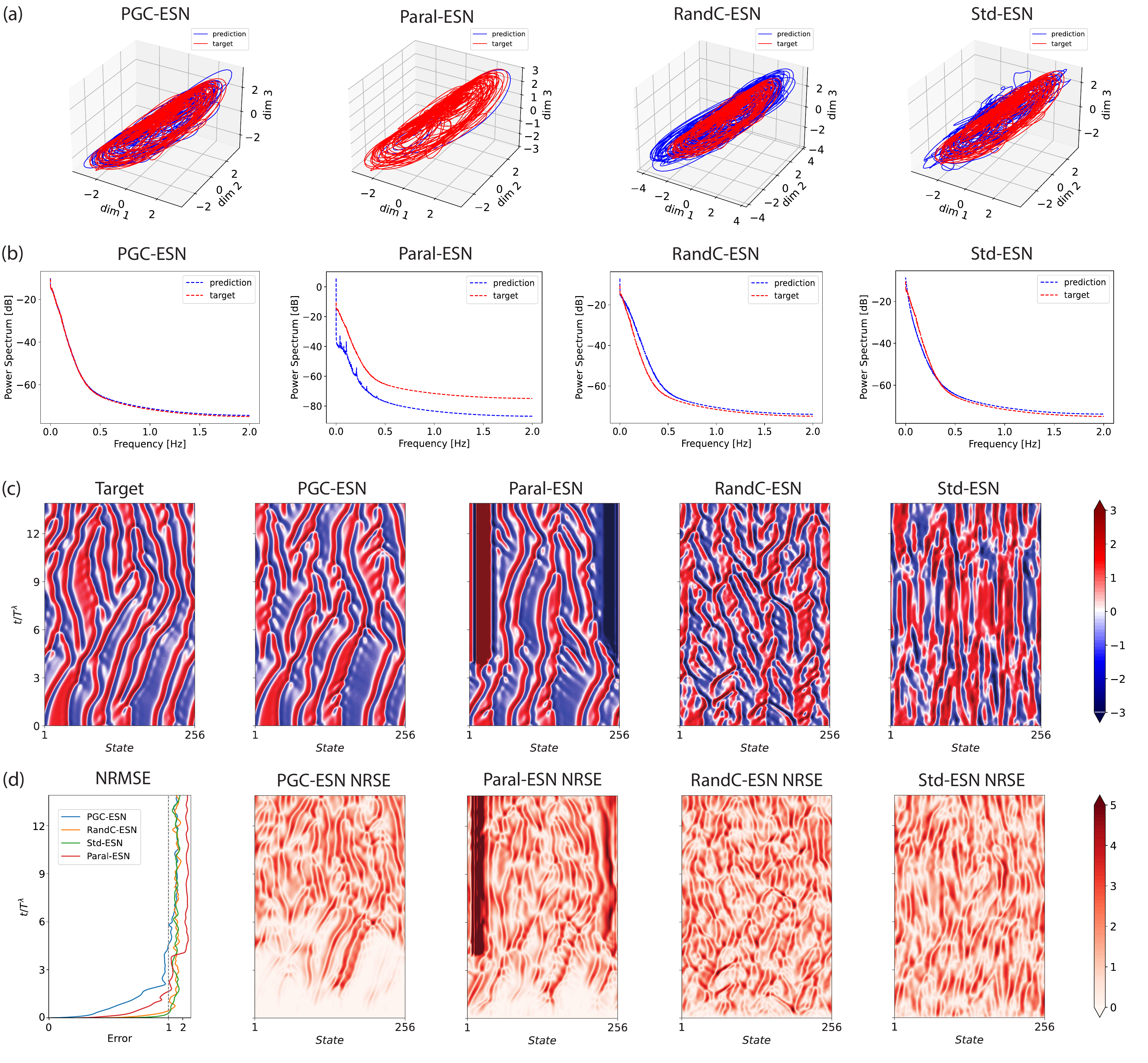}  
    \caption{Predictive results for the inhomogeneous KS equation using a reservoir size of 2000 for PGC-ESN, Paral-ESN, RandC-ESN, and Std-ESN (from left to right). (a) The attractor on the first three-dimensional subspace. (b) The power spectrum. (c) Contour plots of the target and predicted sequences (d) NRMSE evolution and contour plots of the NRSE evolution.}
    \label{fig:inhomo_KS_reservoir_size_2000_plots}
\end{figure*}

Fig.~\ref{fig:homo_KS_reservoir_size_2000_plots} presents the reconstruction results for a specific target sequence of the homogeneous Kuramoto–Sivashinsky (KS) equation using a reservoir size of 2000 with the PGC-ESN, Paral-ESN, RandC-ESN, and Std-ESN (from left to right).
Subplots (a) and (b) depict attractor reconstruction on the first three-dimensional subspace and power spectrum reconstructions, respectively. These results indicate that at a reservoir size of 2000, PGC-ESN achieves the best long-term attractor reconstruction for this sequence. Subplot (c) further reveals that PGC-ESN provides the best short-term predictive power.
Fig.~\ref{fig:inhomo_KS_reservoir_size_2000_plots} presents the results for the inhomogeneous KS equation under the same settings, indicating similar findings.
In conclusion, our proposed PGC-ESN at reservoir size 2000 demonstrates strong capabilities for both long-term attractor reconstruction and short-term prediction with both the homogeneous and inhomogeneous KS equations.

\nocite{*}
\bibliography{paper}

\end{document}